\documentclass[a4paper,onecolumn,11pt,accepted=2022-01-10]{quantumarticle}
\pdfoutput=1

\usepackage[caption=false]{subfig}

\usepackage[square,numbers]{natbib}
\usepackage{graphicx}
\usepackage{amsmath}
\usepackage{amssymb}
\usepackage{amsfonts}
\usepackage{bbm}
\usepackage{braket}
\usepackage{verbatim}
\usepackage{hyperref}
\usepackage{ulem}
\hypersetup{
  colorlinks   = true, 
  urlcolor     = blue, 
  linkcolor    = blue, 
  citecolor   = blue 
}
\usepackage{wrapfig}

\usepackage{color}
\usepackage{tikz}

\newcommand{\PP}{\mathcal P}

\newcommand{\be}{\begin{equation}}
\newcommand{\ee}{\end{equation}}

\newcommand{\tr}[1]{\text{Tr}\big[{#1}\big]}

\begin{document}
	
	\title{Counterdiabaticity and the quantum approximate optimization algorithm}
	
	\author{Jonathan Wurtz}
	\email[Corresponding author: ] {jonathan.wurtz@gmail.com}
	\author{Peter Love}
	\affiliation{Department of Physics and Astronomy, Tufts University, Medford, Massachusetts 02155, USA}


	\begin{abstract}

	The quantum approximate optimization algorithm (QAOA) is a near-term hybrid algorithm intended to solve combinatorial optimization problems, such as MaxCut. QAOA can be made to mimic an adiabatic schedule, and in the $p\to\infty$ limit the final state is an exact maximal eigenstate in accordance with the adiabatic theorem. In this work, the connection between QAOA and adiabaticity is made explicit by inspecting the regime of $p$ large but finite. By connecting QAOA to counterdiabatic (CD) evolution, we construct CD-QAOA angles which mimic a counterdiabatic schedule by matching Trotter ``error" terms to approximate adiabatic gauge potentials which suppress diabatic excitations arising from finite ramp speed. In our construction, these ``error" terms are helpful, not detrimental, to QAOA. Using this matching to link QAOA with quantum adiabatic algorithms (QAA), we show that the approximation ratio converges to one at least as $1-C(p)\sim 1/p^{\mu}$. We show that transfer of parameters between graphs, and interpolating angles for $p+1$ given $p$ are both natural byproducts of CD-QAOA matching. Optimization of CD-QAOA angles is equivalent to optimizing a continuous adiabatic schedule. Finally, we show that, using a property of variational adiabatic gauge potentials, QAOA is at least counterdiabatic, not just adiabatic, and has better performance than finite time adiabatic evolution. We demonstrate the method on three examples: a 2 level system, an Ising chain, and the MaxCut problem.
	\end{abstract}

	\maketitle
	
	\section{Introduction}

	Recently, the quantum approximate optimization algorithm (QAOA) \cite{farhi2014quantum} has grown in interest as a possible algorithm to apply on noisy intermediate scale quantum (NISQ) devices \cite{Preskill_2018,bharti2021noisy}. Given a parameterized wavefunction, a classical optimizer maximizes the expectation value of some objective function, which encodes approximate solutions to \text{NP-HARD} combinatorial optimization problems such as MaxCut. The ansatz alternates between unitaries generated by a ``simple" $H_S$ and ``target" $H_T$ Hamiltonian, which act $p$ times on the maximal eigenstate of the simple Hamiltonian. The ansatz wavefunction, parameterized by $2p$ variational angles, $\gamma,\beta$ is
	
	\begin{equation}\label{eq:THE_QAOA_ANSATZ}
	    |\gamma,\beta\rangle \;=\; \exp(i\beta_p H_S)\exp(i\gamma_p H_T)\;(\cdots)\;\exp(i\beta_1 H_S)\exp(i\gamma_1 H_T)\;|0_S\rangle.
	\end{equation}

	In the original introduction of QAOA, Farhi et.~al.~\cite{farhi2014quantum} show that in the limit $p\to\infty$, the QAOA is exact, in that the ansatz sate can be in the subspace of the maximal eigenstates of the objective function. Their proof leverages the adiabatic theorem: Given a very slowly time varying Hamiltonian, an initial eigenstate will evolve as an instantaneous eigenstate of the time varying Hamiltonian. Under adiabatic evolution, a simple-to-prepare ground state of an initial Hamiltonian is ``dragged along" to prepare a ground state of a final Hamiltonian that encodes the solution to a combinatorial optimization problem \cite{Lucas2014,Albash2018}. The adiabatic condition requires that the duration of the evolution be (roughly) bounded by the inverse gap squared \cite{TEST}. In the $p\to\infty$ limit, a set of perturbatively small QAOA angles can satisfy the adiabatic condition by reproducing a Trotterized adiabatic evolution. A na\"ive choice of angles \cite{sack2021quantum} are angles incrementally increasing and decreasing along each step $q$ as $\gamma_q=\gamma_0 (q/p)$ and $\beta_q = \beta_0(1-q/p)$. If the angles are perturbatively small $\gamma_0,\beta_0 \ll 1$, the Trotter error is suppressed and the $q$th QAOA step reproduces a continuous time unitary for a time step $\tau=|\gamma_0| + |\beta_0|$ along an effective QAOA Hamiltonian
	
	\begin{equation}
	    H_q = (q/p) H_T \;+\; (1 - q/p) H_S \;-\; \frac{i(q/p)(1-q/p)\gamma_0\beta_0}{2(\gamma_0+\beta_0)}\big[H_T,H_S\big] \;+\;\cdots
	\end{equation}
	where ellipsis denote higher order terms of a Baker-Campbell-Hausdorff (BCH) expansion. In the limit $\gamma_0,\beta_0\to 0$, the higher-order BCH terms are zero, and the effective Hamiltonian reproduces a linear annealing schedule between simple and target Hamiltonian. In the limit $p\to\infty$, the total time $T\to\infty$, which satisfies the adiabatic condition and so the final QAOA ansatz state is an exact maximal state.
	
	\quad
	
	The $p\to\infty$ limit is excessive. It is the limit in which the angles for each step are perturbatively small to minimize the Trotter error, while the sum of all angles (eg.~the total time) is large to enforce the adiabatic condition. To satisfy both, $p$ must be very large. What happens when $p$ is large but finite? This paper focuses on relaxing the $p\to\infty$ condition, computing and simulating QAOA in the large but fixed-$p$ regime. It will be shown that QAOA angles at finite $p$ can reproduce, and perform better than, a finite time adiabatic evolution. 
	\begin{figure}
        \centering
        \includegraphics{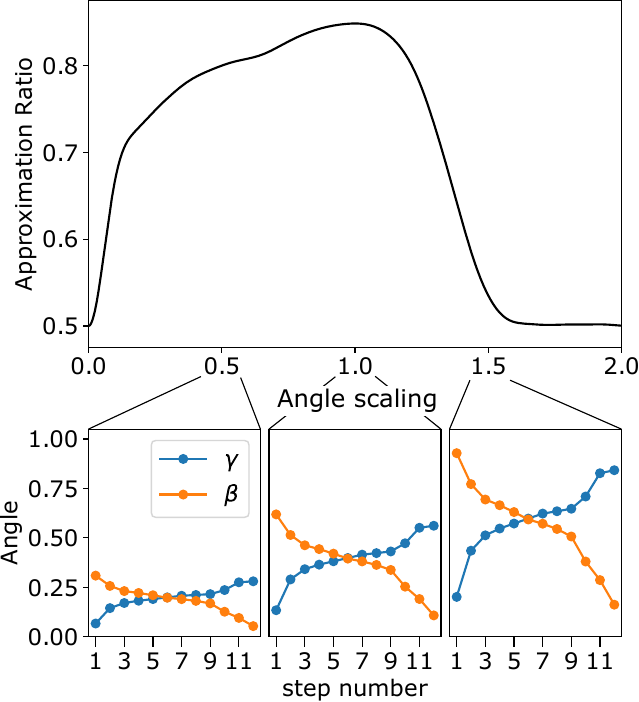}
        \caption{Optimizing CD-QAOA angles of fixed depth $p$. Naively, a smooth set of angles that mimic an adiabatic process leaves the relative magnitude of each angle undefined. If the angles are too small (left), the adiabatic timescale is too short and the state accumulates an excess of finite speed diabatic excitations, reducing the performance. If the angles are too large (right), the Trotter error of alternating between Hamiltonian terms accumulates, reducing the performance. Somewhere in between, these two error sources accumulate minimally, resulting in an optimum. We seek this optimum by matching the Trotter error with low order counterdiabatic terms. These data are variationally optimized $p=12$ angles for a MaxCut graph, as further described in Fig.~\ref{fig:3r_derived_protocol}.}
        \label{fig:angle_scaling_sketch}
    \end{figure}
    The main result of this work is that the Trotter ``error" from alternating between terms in the generator can be a boon, not a bane, for a finite-$p$ QAOA that mimics adiabatic protocols, by suppressing diabatic excitations. This leads to the fact that QAOA is at least \textit{counterdiabatic}, not just \textit{adiabatic}.
    
    The Trotter error, to lowest order, gives a contribution to the effective QAOA Hamiltonian which is the commutator of the simple and target Hamiltonian $[H_T,H_S]$, where the magnitude of the angles $\gamma$, $\beta$ give the relative strength of the Trotter error term. There is a trade off in the magnitude of these angles, as sketched in Fig.~\ref{fig:angle_scaling_sketch} for fixed $p$. If the angles are too small (Fig.~\ref{fig:angle_scaling_sketch} left), the adiabatic timescale is too short, and the performance is lessened by finite-speed diabatic excitations. If the angles are too large (Fig.~\ref{fig:angle_scaling_sketch} right), the Trotter error from combining digital terms into a combined unitary is large, accumulating error and reducing the performance. Somewhere in the middle (Fig.~\ref{fig:angle_scaling_sketch} center), these two error sources accumulate error minimally, resulting in a maximum performance.
    
    This maximum can be found analytically by matching the Trotter ``error" term to a low-order approximation of the adiabatic gauge potential (AGP), and thus show that the first-order Trotter error can act as a diabatic counter term to reduce excitations. The AGP and the broader concept of ``Shortcuts to adiabaticity" (STA) \cite{Guery2019} covered in section \ref{sec:shortcuts_to_adiabaticity} aims to reduce finite-speed diabatic excitations \cite{Rigolin2008,Bachmann2017} by adding counter terms to the effective Hamiltonian. In doing so, the magnitude of QAOA angles is fixed by an analytically computed coefficient related to the variational AGP, and the total set of angles can be analytically computed. A counterdiabatic (CD) or STA Hamiltonian has better finite-time performance over analogous adiabatic-only protocols, and so these analytically-derived QAOA angles have better performance than an analogous adiabatic-only protocol.
    
    Crucially, we show that the second order BCH term can always be matched to a low order variational adiabatic gauge potential, as both the sign of the BCH term and the AGP coefficient are negative. The implication of this matching is that for counterdiabatic annealing protocols of total time $T$, there exists some finite $p(T)$ which is optimal. The optimal $p(T)$ connects performance scaling of adiabatic processes with QAOA scaling with $p$.

    \quad
    
    To demonstrate using the Trotter error as a counterdiabatic term, we construct an explicit procedure, which maps between continuous time counterdiabatic and STA protocols and discrete QAOA angles. The procedure is completely analytic, numerically implementable, and does not require quantum simulation. The forward procedure, given a counterdiabatic Hamiltonian $H_\text{CD}(t)$, is outlined as follows. Each QAOA step is expanded into a single unitary using a perturbative Baker-Campbell-Hausdorff (BCH) expansion, which is reasonably numerically tractable for perturbation order $\lesssim 5$. The generator of the step is matched with the effective counterdiabatic time evolution generator over some finite step in time, computed using a low order ($\lesssim 5$) Magnus expansion. The effective generators along each step $q$ of QAOA are matched with the time evolution generators along some time step by classically variationally optimizing QAOA angles $\gamma_q,\beta_q$ as well as time step $\tau_q$ to minimize some trace error. If the matching is close for every step, the QAOA ansatz state is close to the time evolved counterdiabatic state. If the first-order counterdiabatic term is exact, as is the case for the two level system, then an exact $p=1$ QAOA can be derived using these methods, as shown in section \ref{sec:2level_system}.
    
    Matching QAOA angles to finite-speed adiabatic schedules has several interesting corollaries. One useful implication is deriving the large-$p$ behavior as the approximation ratio converges to 1. Kibble Zurek (KZ) theory \cite{del_Campo_2014} relates energy density to finite-speed annealing through quantum critical points \cite{Latorre2004}. Using KZ scaling we show that the large-$p$ QAOA behavior of the approximation ratio approaches unity at least polynomially
    
    \begin{equation}
        C(p) \sim 1 - p^{-\mu}.
    \end{equation}
    
    Curiously, such scaling may be slower than optimal QAOA angles for a given $p$. We show in section \ref{sec:ising_chain} that $C(p)\sim 1-p^{-1/2}$ for the Ising chain and finite-speed protocols. For optimal protocols \cite{Barankov2008}, we show that $C(p)\sim 1-p^{-1}\log(p)$. However, the optimal QAOA angles \cite{farhi2014quantum} go as $C(p) = 1 - 1/2p$. This scaling suggests that QAOA angles which mimic counterdiabatic evolution may not be optimal in all systems, and globally optimal protocols may be hard to find.

    The map between continuous time counterdiabatic protocols and QAOA angles can be done in reverse. Given a set of smooth QAOA angles, an approximately equivalent continuous time counterdiabatic protocol with varying speeds $\lambda(t)$ and auxiliary fields $s(t)$ can be derived by variationally optimizing or matching each QAOA step with an equivalent continuous-time counterdiabatic Hamiltonian. This procedure will be shown in section \ref{sec:METHODS} and demonstrated in section \ref{sec:3regular_graphs}. Consequently, variational optimization of smooth QAOA angles in a large $p$ limit is equivalent to variational optimization of continuous time counterdiabatic protocols \cite{schiffer2021adiabatic}. The phenomena of  inducing larger-$p$ protocols \cite{Zhou_2020} is also naturally explained by this mapping. Given an induced counterdiabatic protocol for some $p$, one can compute new angles for a different $p'$, which will have a similar ``shape", and similar performance.

    Crucial to the matching procedure is the addition of some ancillary field $s$, which adds extra terms to the effective QAOA Hamiltonian
    
    \begin{equation}
        H(t)\;\mapsto\; H(t) + s(t) \tilde H(t).
    \end{equation}
    
    Here, $\tilde H(t)$ slowly changes and includes terms from the BCH expansion, and $s(t)\to0$ at the beginning and end of the protocol, so that in the $T\to\infty$ the final state is the exact ground state of the target Hamiltonian. This implies that that the QAOA state, instead of following the ground state of the original Hamiltonian $\lambda H_S+(1-\lambda)H_T$, follows the ground state of some other Hamiltonian which follows a more complicated path in a multi-dimensional parameter space. In this sense, matching QAOA with a simple Trotterized adiabatic path is inadequate to describe the effective evolution, and these extra terms are required.
    
    Additionally, if two graphs have a similar low energy eigenspectrum, it is reasonable to expect that the induced optimal counterdiabatic annealing protocol for the first graph $\mathcal G$ will have similar performance for the other graph $\mathcal G'$. Consequentially, the induced angles for a graph $\mathcal G'$ given the counterdiabatic protocol induced from variationally optimized angles from graph $\mathcal G$ will have similar performance. This counterdiabatic insight may explain the phenomena of transfer of parameters \cite{brandao2018fixed}, where optimal angles for one graph are close to optimal for another. A more in-depth discussion of transfer of parameters in the counterdiabatic context is shown in sections \ref{sec:METHODS} and \ref{sec:3regular_graphs}.
    
    Finally, we show that, in a certain sense, QAOA does better than an analogous continuous time adiabatic evolution. Given a total ``angle budget" $T=\sum_q|\gamma_q|+|\beta_q|$, a set of QAOA angles which mimic an adiabatic schedule does better than a finite-time smooth evolution of total time $T$. This fact is a consequence of Pontryagin’s minimum principle \cite{Yang_2017,Brady2021}. We show that more specifically, this is due to the fact that the sign of the variational adiabatic gauge potential is always the same as the first-order Trotter ``error" term. This makes the effective QAOA evolution at least counterdiabatic, not just adiabatic, which suppresses excitations and improves performance. This fact is discussed further in Sec.~\ref{sec:QAOA_is_better_than_adiabatic}. 
    
    We comment that while this mapping is perturbatively exact in the large $p$ limit, the method nonetheless produces protocols with good performance even for small $p\sim 1$. For instance, the method finds the exact $p=1$ angles for a 2 level system, as shown in section \ref{sec:2level_system}, and comes very close to the exact angles for the Ising chain, as shown in section \ref{sec:ising_chain}. These observations indicate that analytically computed CD-QAOA angles may be pre-computed without any variational feedback step \cite{Streif_2020} and used directly, even for the low depths $p\sim 5$ of tomorrow's NISQ hardware.
    
    \section{Outline of the paper}
    
    The rest of the paper is as follows. Section \ref{sec:shortcuts_to_adiabaticity} begins by introducing the adiabatic theorem, adiabatic protocols, and the concept of shortcuts to adiabaticity (STA). This section will also introduce the variational adiabatic gauge potential, which will be matched with the Trotter ``error" terms. Section \ref{sec:METHODS} introduces the method of matching continuous time counterdiabatic evolution with QAOA angles, as well as justifies the perturbative expansions of the matching procedure in the large $p$ limit. Then, sections \ref{sec:2level_system}-\ref{sec:3regular_graphs} demonstrates the methods on three examples. Section \ref{sec:2level_system} outlines the method on a simple two level system, and show that the derived QAOA angles are exact for all $p\geq 1$ up to perturbative error, as expected. Section \ref{sec:ising_chain} outlines the method on the 1d Ising chain, where the variational AGP is no longer exact but the model is exactly solvable. Section \ref{sec:3regular_graphs} outlines the method on 3 regular graphs, where the goal is to find approximate solutions to MaxCut. This section will also illustrate reversing the CD-QAOA procedure to derive approximately optimal counterdiabatic protocols. Finally, sections \ref{sec:QAOA_is_better_than_adiabatic}-\ref{sec:conclusion} concludes with some discussion of the implications of CD-QAOA, and some interesting next steps for the method.

	\quad

	\section{Adiabatic protocols and shortcuts to adiabaticity}\label{sec:shortcuts_to_adiabaticity}
	
	An adiabatic protocol is generally a process which leave certain dynamical properties invariant \cite{Born1928}. For example, under time evolution which slowly changes the underlying Hamiltonian and eigensystem, an adiabatic process could leave the relative probability distribution of eigenstates of the time-evolved wavefunction unchanged. For this reason, a common implementation of adiabatic protocols is state preparation and quantum adiabatic algorithms (QAA) \cite{farhi2000,Albash2018}. Given some simple to prepare state $|0_S\rangle$ which is an eigenstate (usually ground or maximal) of a Hamiltonian $H_S$, parameters are slowly tuned to turn off $H_S$ and turn on some target Hamiltonian $H_T$, where the (ground or maximal) eigenstate is the target state $|0_T\rangle$. A particular time dependent protocol which achieves this simply interpolates between the two with some time dependent fields $\lambda(t)$ and $s(t)$, such that $\lambda(0)=0$, $\lambda(T)=1$, and $s(0)=s(T)=0$

	\begin{equation}\label{eq:adiabatic_only_hamiltonian}
	    H(t) = \lambda(t) H_T + \big(1-\lambda(t)\big) H_S + s(t) \tilde H(t).
	\end{equation}
	
	Here, we also include some slowly changing auxiliary field $\tilde H(t)$ which may bypass gap closings to speed adiabatic annealing. If the rate of change of the Hamiltonian as a function of time is small in comparison to the energy gap between the eigenstate and neighboring states, the initial eigenstate is ``dragged along" and persists as an eigenstate. The speed is characterized by the adiabatic condition $1/\dot\lambda \gg |\langle 0|\partial_t H|n\rangle| / \Delta_n^2$, where $n$ indexes any other instantaneous eigenstate, and $\Delta_n$ is the energy difference \cite{TEST}. If the condition is violated, the eigenstate accumulates diabatic excitations which, undesirably, makes the state no longer an eigenstate. This condition can occur especially around phase transitions, where the gap becomes very small and diabatic excitations are unavoidable for large system sizes through processes such as the Kibble-Zurek mechanism (KZ) \cite{del_Campo_2014}.

	The field is usually taken to be a linear ramp $\lambda(t)=t/T$ for a total time $T$. Given a finite time to anneal between simple and target Hamiltonians a more general nonlinear field may be more advantageous, which is the goal of optimal control theory \cite{Rezakhani2009,Brif_2014,Brady2021,brady2021behavior}. Without constraint on the smoothness of the protocol, optimal protocols are ``Bang-Bang" QAOA protocols in accordance with Pontryagin’s minimum principle \cite{Yang_2017} as long as the controls are nonsingular \cite{Seraph2018}. Similar to QAOA, good protocols may include a smooth ramp, plus a rapidly oscillatory term \cite{Chasseur2015,Claeys2019}. Depending on the structure of the gap, an optimal smooth protocol may go fast where the gap is large, go slowly around phase transitions, where the gap is small \cite{Barankov2008,Sachdev2011a}. In this work we focus on second order phase transitions, characterized by smooth changes in the ground state. This is opposed to first order phase transitions, which are characterized by sharp changes in order parameter and exponentially small level crossings. However, it may be possible to change from a first to second order phase transition by eg.~adding non-stoquastic terms to a stoquastic Hamiltonian \cite{Nishimori2017}.
	
	Quantum computers are usually limited in the total time for annealing \cite{Johnson2011} and so it is advantageous to find fast protocols which add extra terms to the time dependence to avoid diabatic excitations or generally get a high-quality final state $|0_T\rangle$. This is the goal of Shortcuts to adiabaticity (STA) \cite{Guery2019}: given control parameters $\lambda(t)$ and extra auxiliary fields $s(t)$, choose specific time dependent protocols which optimize the final state of the process, which may include non-adiabatic intermediary processes.
	
	We will use two specific concepts from STA to construct QAOA angles. The first concept is path dependent protocols \cite{Brif_2014,Sugira2021} 
	which choose nonzero values of auxiliary fields to find optimal paths between initial and final Hamiltonians in the multidimensional parameter space of control fields, potentially avoiding regions where diabatic excitations are likely. The second concept is counterdiabatic protocols \cite{Berry_2009}, which add counter terms to the Hamiltonian to suppress diabatic excitations which occur from the finite speed evolution. Formally, the exact counter term is the adiabatic gauge potential (AGP) \cite{KOLODRUBETZ20171}, which serves as the generator of parameter evolution of instantaneous eigenstates. For typical Hamiltonians, the AGP is exponentially large in norm and highly nonlocal, rendering any experimental implementation impossible \cite{Pandey2020}.
    For this reason, implementable counterdiabatic protocols must use approximate AGPs which are local. One choice are variational AGPs \cite{Sels2017}, which use some local parameterized ansatz $A(\alpha)$ for an approximation of the AGP and best match it with the exact AGP.
    This is done though a trace minimization of an action
	
	\begin{equation}\label{eq:AGP_trace_minimization}
	    \underset{\alpha}{\texttt{ MIN:}}\quad S(\alpha) = \Big|\Big|\partial_\lambda H - i[A(\alpha),H]\Big|\Big|_{\mathcal P},
	\end{equation}
	which suppresses the diabatic excitations of each eigenstate, averaged over the positive semidefinite projector $\mathcal P\succcurlyeq 0$ \cite{KOLODRUBETZ20171}. The norm is defined as $||Q||_\PP = \tr{\mathcal P Q^2}$. Note that if $\mathcal P=\mathbbm 1$ this trace minimization is analytically tractable for any number of qubits, using trace identities to avoid any computation in Hilbert space. Recently, there has been a proposal for an intuition-agnostic ansatz for the AGP as an expansion of commutators \cite{Claeys2019}
	
	\begin{equation}\label{eq:claeys_expansion}
	    A(\alpha) \;= \;i\alpha_0 [H,\partial_\lambda H] \;+\; i\alpha_1 [H,[H,[H,\partial_\lambda H]]] \;+\; i\alpha_n[H,[\cdots,[H,\partial_\lambda H]\cdots ]] \;+\; \dots
	\end{equation}
	which is an expansion of Hasting's formulation of the adiabatic gauge potential for adiabatic continuation \cite{Hastings2005}. Using the Hamiltonian of Eq.~\eqref{eq:adiabatic_only_hamiltonian}, the lowest-order AGP from Eq.~\eqref{eq:claeys_expansion} is simply the commutator of the simple and target Hamiltonian $A = i\alpha_0 [H_S,H_T]$. A critical observation that will be used repeatedly though this work is the fact that the lowest order AGP ansatz is equivalent to the lowest order of BCH expansion used to compute the effective generator of each QAOA step. The coefficient $\alpha$ of Eq.~\eqref{eq:AGP_trace_minimization} can be optimized by setting $\partial_\alpha S(\alpha) =0$. Assuming that the projector $\PP$ is diagonal in the instantaneous eigenbasis of $H$, using some trace identities the coefficient of the variational AGP is

	\begin{equation}\label{eq:alpha_definition}
	    A(\lambda) = i\alpha(\lambda) [H_T,H_S]\quad\text{with}\quad \alpha(\lambda) = \frac{\Big|\Big|[H,\partial_\lambda H]\Big|\Big|_\PP}{\Big|\Big|\big[[H,\partial_\lambda H],H\big]\Big|\Big|_\PP}.
	\end{equation}
	
	Observe that the sign of $\alpha$ is always negative over all projectors $\PP\succcurlyeq 0$. The numerator is the norm of an antihermitian operator, which is negative; the denominator is the norm of a Hermitian operator, which is positive. The sign of $\alpha\leq 0$ even if $\PP = |0\rangle\langle 0|$, the instantaneous ground state, or any low-energy subspace. The norm only includes at most the sixth power of the Hamiltonian and so the AGP is only aware of local structure of the system, albeit over a larger neighborhood than the Hamiltonian terms.
	
	The approximate AGP can be added to the Hamiltonian as a counter term with a strength proportional to the speed $\dot\lambda$ \cite{KOLODRUBETZ20171}. Additionally, we add an auxiliary field $s$ that is also the same commutator term. The effective time evolution operator (eg, counterdiabatic Hamiltonian) becomes
	
	\begin{equation}\label{eq:counterdiabatic_hamiltonian}
	    H_\text{CD}(\lambda,\dot\lambda,s) = \lambda H_T \;+\; \big(1-\lambda\big) H_S \;+\; i\big( s + \dot \lambda \alpha(\lambda)\big)\big[H_T,H_S\big].
	\end{equation}
	
	In the limit $T\to 0$, the counterdiabatic Hamiltonian is equivalent to evolving in parameter space $\lambda\in 0\to 1$ with respect to the variational AGP, and in the limit $T\to\infty$ and $s=0$ Eq.~\eqref{eq:counterdiabatic_hamiltonian} recovers the original adiabatic schedule of Eq.~\eqref{eq:adiabatic_only_hamiltonian}. It is a physically reasonable conjecture to expect
	that adding an approximate adiabatic gauge potential to the Hamiltonian as a counter term will decrease the amount of diabatic excitations to extremal states. Even though the variational parameter is optimized to minimize excitations over all eigenstates (Eq.~\eqref{eq:AGP_trace_minimization}), it has been found that the AGP still works in the $T\to 0$ limit for extremal eigenstates of interacting models \cite{Wurtz2020b}.
	Similarly, it is reasonable to expect that a local approximation of the AGP will suppress transitions far apart in energy \cite{Hastings2005}.
	and that a local AGP will generally mimic the action of the full AGP away from phase transitions. Reducing diabatic excitations will minimize energy variance and thus increase the value of the objective function (for optimization) or decrease energy density (for ground states), resulting in better performance of the algorithm.
	
	Note that this reasonable conjecture may be false in certain cases. If the AGP is optimized for all eigenstates, and it may be the case that optimizing for low energy or ground states only may result in a smaller value of $\alpha$. Given an extreme enough difference, a counterdiabatic evolution may ``overcorrect" and result in negative performance for the ground state, while having on average positive performance over all eigenstates. A characterization of such systems is beyond the scope of this paper. Nonetheless, the value of $\alpha\leq 0$ even for the ground state, so an infinitesimal value of $\alpha$ will always improve performance.
	
	\section{Counterdiabatic QAOA}\label{sec:METHODS}
	
	\begin{figure}
	    \centering
	    \includegraphics[width=\linewidth]{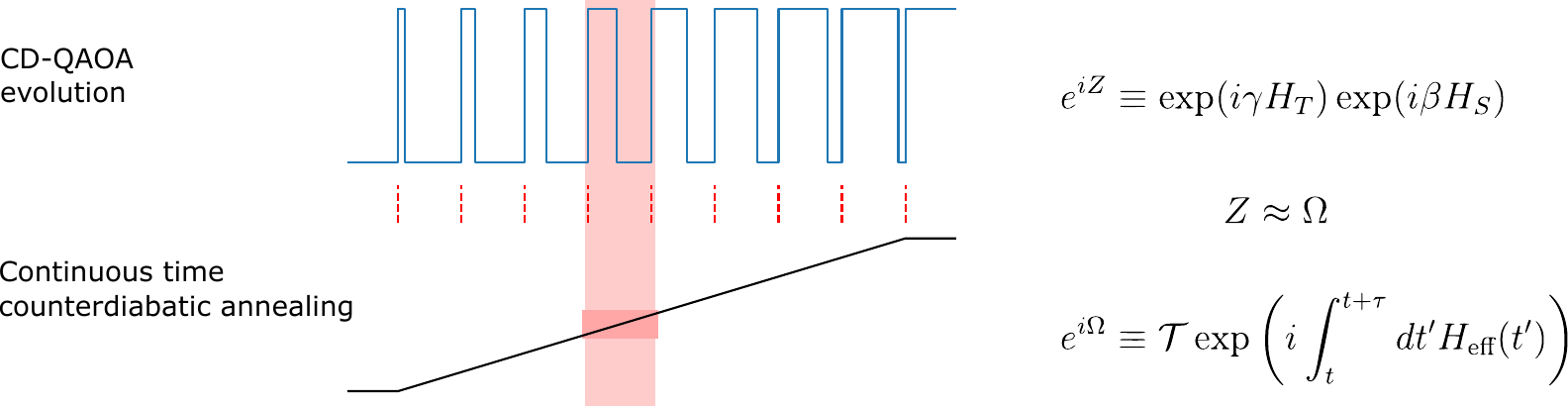}
	    \caption{Schematic diagram of the CD-QAOA angle matching procedure. As input, the procedure receives the counterdiabatic annealing protocol (Black line), effective Hamiltonian $H_\text{eff}$, and number of QAOA steps $p$. For each step of the CD-QAOA evolution (pink), the effective unitaries over the interval are matched by minimizing the difference of their generators (right). This procedure sets the value of each angle $\gamma_q$, $\beta_q$ in the QAOA (blue). By best matching along each step, the procedure fixes the angles $\{\gamma,\beta\}$ and total adiabatic annealing time $T$.}
	    \label{fig:QAOA_matching_schematic}
	\end{figure}
	
	We will now derive the mapping between continuous time counterdiabatic annealing, and QAOA angles. The QAOA ansatz of Eq.~\eqref{eq:THE_QAOA_ANSATZ} is a set of unitaries that alternates between the simple Hamiltonian and the target Hamiltonian, which act on the maximal eigenstate of the simple Hamiltonian. In general, the $2p$ optimal angles $\{\gamma,\beta\}$ may be arbitrary, with the difference of angles between steps being large. A subset of QAOA angles are those which are smooth, where the difference of angles between steps is small. Such QAOA angles mimic an adiabatic process by smoothly interpolating between $\beta$ large $\gamma$ small at the beginning, and $\beta$ small $\gamma$ large at the end. Smooth QAOA angles are herein defined as a \emph{counterdiabatic QAOA} (CD-QAOA). A set of CD-QAOA angles is identified by best matching the unitary generated by one step of QAOA, with the unitary generated by time evolution with the counterdiabatic Hamiltonian between some time interval $[t,t+\tau]$, for each QAOA step $q$
	
	\begin{equation}\label{eq:BCH_magnus_matching}
	    \exp(i\gamma H_T)\exp(i\beta H_S)\equiv e^{iZ}\approx e^{i\Omega}\equiv \mathcal T\exp\bigg(i\int_t^{t+\tau}dt' H_\text{eff}(t')\bigg).
	\end{equation}
	
	A schematic of this matching is shown in Fig.~\ref{fig:QAOA_matching_schematic}. $Z$ and $\Omega$ are the generators of each unitary, which may be defined perturbatively in a Baker-Campbell-Hausdorff (BCH) expansion \cite{Hatano_2005} and Magnus expansion \cite{Blanes_2009}, respectively.
	
	To lowest order of small time steps $\tau\ll 1$, the values of $\gamma$ and $\beta$ can be computed by de-Trotterizing the QAOA step by computing $Z$ to the lowest order in BCH expansion, and computing $\Omega$ to the lowest order in the Magnus expansion. The lowest order BCH is simply a sum of the terms, and the lowest order Magnus expansion is simply the time average of the time dependent adiabatic-only Hamiltonian. For some small time step $\tau\ll1$ the generators are matched by fixing angles
	
	\begin{align}\label{eq:simplest_adiabatic_QAOA}
	    Z&\approx \Omega,\\
	    \gamma H_T + \beta H_S &\approx \tau\lambda(t)H_T + \tau(1-\lambda(t))H_S;\\
	    &\Rightarrow\quad\gamma = \tau \lambda(t).\\
	    &\Rightarrow\quad\beta = \tau (1-\lambda(t)).
	\end{align}
	
	The Magnus and BCH expansion, as well as matching generators instead of unitaries, works in the perturbative limit $\tau\to 0$. For sufficiently large $T$ to suppress diabatic excitations from finite-time adiabatic adiabatic evolution, and sufficiently small $\tau$ to suppress Trotterization errors, one can construct an adiabatic QAOA protocol with $p=T/\tau$ steps which asymptotically returns the ground state of $H_T$. This is the argument of Farhi et al.~in the original introduction of QAOA \cite{farhi2014quantum}. However, this protocol leaves the relative magnitude of angles $\{\gamma,\beta\}$ free, and requires the limit $\tau\to0$ to suppress higher order BCH error.

	The angles can be fixed, and BCH error can be used advantageously, by using the counterdiabatic insights developed in section \ref{sec:shortcuts_to_adiabaticity}. Given some fixed finite $p$, one can still construct approximately adiabatic QAOA parameters as follows. First, instead of using the adiabatic-only Hamiltonian of Eq.~\eqref{eq:adiabatic_only_hamiltonian}, one can use the counterdiabatic Hamiltonian of Eq.~\eqref{eq:counterdiabatic_hamiltonian} as the generator of time evolution. To better match generators, one can extend the effective QAOA evolution generator to the next order in the BCH expansion. Angles can then be fixed by matching the effective generators; observe that, crucially, the second order BCH term is the same as the approximate AGP and auxiliary term
	
	\begin{align}
	    Z\approx&\;\Omega,\\
	    \gamma H_T + \beta H_S - \frac{i\gamma\beta}{2}[H_T,H_S]\approx&\;
	    \tau \overline \lambda H_T  + \tau (1-\overline \lambda ) H_S +i\tau\big( \overline s + \dot\lambda \overline{\alpha}\big)[H_T,H_S];\label{eq:1st_order_matching}\\
	    \Rightarrow &\quad\gamma=\tau\overline \lambda,\label{eq:gamma_forced}\\
	    \Rightarrow &\quad\beta=\tau(1-\overline \lambda),\label{eq:beta_forced}\\
	    \Rightarrow &\quad\frac{-\gamma\beta}{2}=\tau\bigg(\overline s + \dot\lambda \overline \alpha\bigg).\label{eq:tau_forced1}
	\end{align}
	
	Matching the second order BCH expansion to the CD Hamiltonian sets not only the angles to fixed values, but also the time step
	
	\begin{align}\label{eq:lowest_order_gamma_beta_fixing}
	    \tau = \frac{-2(\overline s +\dot\lambda\overline\alpha )}{\overline \lambda(1-\overline\lambda)}\quad;\quad \gamma = \frac{-2(\overline s +\dot\lambda\overline\alpha )}{1-\overline \lambda}\quad;\quad \beta = \frac{-2(\overline s +\dot\lambda\overline\alpha )}{\overline\lambda}.
	\end{align}
	
	Note that $\alpha<0$ by Eq.~\eqref{eq:alpha_definition}, and so this solution always exists, as long as $\overline s$ is not to large. In order for the expansions to be controlled, the next order terms must be much smaller. For the BCH expansion the perturbative criterion is that the third order term (see Eq.~\eqref{eq:BCH_expansion}) is much smaller than the second
	
	\begin{equation}
	    \bigg|\frac{\gamma^2\beta}{12}\bigg| + \bigg|\frac{\beta^2\gamma}{12}\bigg|\ll \bigg|\frac{\gamma\beta}{2}\bigg|\quad \Rightarrow\quad \bigg|\frac{s+\dot\lambda \alpha}{3(1-\lambda)}\bigg|+\bigg|\frac{s+\dot\lambda \alpha}{3\lambda}\bigg|\ll1.
	\end{equation}
	
	For the Magnus expansion the perturbative criterion is that the second order term (See Eq.~\eqref{eq:magnus2}) is much smaller than the first
	
	\begin{equation}
	    \frac{\dot\lambda \tau^3}{12}||[H_T,H_S]||\ll \tau(||H_T|| + ||H_S||)\quad\Rightarrow\quad\frac{\dot\lambda ( s + \dot\lambda \alpha)^2}{3\lambda^2(1-\lambda)^2}\ll \frac{||H_T|| + ||H_S||}{||[H_T,H_S]||}.
	\end{equation}
	
	For $T\to\infty$, $s\ll 1$, and away from the edges of the protocol, both criterion are satisfied and the expansions are valid. This link allows relating the total time to $p$. From Eq.~\eqref{eq:lowest_order_gamma_beta_fixing}, the time for each QAOA step $\tau$ goes as $\dot\lambda \alpha$ for $s=0$. Given $\dot\lambda \sim 1/T$, the total time $T=\sum\tau$ is quadratic in $p$. For $s< 0$ and $T\to\infty$, the time for each QAOA step $\tau$ goes as $s$, as $s\gg \dot\lambda \alpha$ and the total time $T=\sum\tau$ is linear in $p$
	
	\begin{equation}\label{eq:p_time_relation}
	    -\alpha T^2\sim  p\qquad;\qquad  -sT\sim  p.
	\end{equation}
	
	This scaling is easily verified with dimensional analysis: $\alpha$ has units of inverse energy squared, and so ($\hbar=1$) time must be quadratic in $p$. $s$ has units of inverse energy, and so time must be linear in $p$. Crucially, the sign of $\alpha$ or $s$ must be negative in order to have a positive value of $p$.
	
	\quad
	
	\subsection{General Counterdiabatic QAOA}
	
	The matching procedure of Eq.~\eqref{eq:1st_order_matching} can be made more general by including arbitrarily high orders of BCH and Magnus expansions. For the $q$th step in the counterdiabatic annealing schedule between times $[t_q,t_q+\tau_q]$, compute the effective generators $Z(\gamma_q,\beta_q)$ and $\Omega(t_q,\tau_q)$ through the BCH and Magnus expansions to appropriately high order. Then, make the equality of Eq.~\eqref{eq:BCH_magnus_matching} as close as possible by minimizing the size of the difference $Z-\Omega$.
	
	\begin{equation}
	    \exp\big(iZ(\gamma_q,\beta_q\big)\approx \exp\big(i\Omega(t_q,\tau_q)\big)\quad\Rightarrow\quad \exp\big(iZ - i\Omega\big)\approx\mathbbm{1}.
	\end{equation}
	
	For each step, there is some error $e_i$ which adds to the effective counterdiabatic time evolution
	
	\begin{equation}
	     e_q(\gamma_q,\beta_q,\tau_q) = \frac{\sqrt{\tr{(Z-\Omega)^2}}}{\tau_i\sqrt{N}}.
	\end{equation}
	
	The effective time evolution Hamiltonian which best mimics the QAOA angles is
	
	\begin{equation}
	    H_\text{eff}(t) = \lambda(t) H_T + \big(1-\lambda(t)\big) H_S + i\big( s(t) + \dot \lambda \alpha(\lambda,s)\big)[H_T,H_S] \;+e(t) H_E(t),
	\end{equation}
	where $e(t)\approx e_q$ for $t\in [t_q,t_q+\tau_q]$ and $H_E\propto Z-\Omega$, with norm $||H_s||=N$ is the extra ``error" term. If $e_q\approx 0$ for each QAOA step then the unitaries mimic the continuous time version, and final state of the CD-QAOA is  that of the counterdiabatic time evolution. To make the effective evolution close to the counterdiabatic evolution, one can minimize the total error
	\begin{equation}\label{eq:counterdiabatic_QAOA_algorithm}
	    \underset{\gamma_q,\beta_q,\tau_q}{\texttt{MIN:}}\quad \sum_q e_q(\gamma_q,\beta_q,\tau_q)
	\end{equation}
	by variationally minimizing $3p$ parameters $\gamma_q,\beta_q,\tau_q$. This minimization completely fixes the CD-QAOA angles $\{\gamma,\beta\}$, as well as optimizes the best total evolution time $T=\sum_q\tau_q$ of the analogous CD annealing for given $p$. The trace can be computed efficiently using Pauli identities eg $\tr{(\sigma_x)^2} = 1$ and $\tr{\sigma_x}=0$. The computation of $Z(\gamma,\beta)$ and $\Omega(t,\tau)$ may be unwieldy for large orders of expansion: the number of terms generically grow exponentially with the expansion order. However, such expansions are reasonably tractable with computational tools using Pauli identities, for low order $\sim 5$. Eq.~\eqref{eq:counterdiabatic_QAOA_algorithm} reduces to Eq.~\eqref{eq:lowest_order_gamma_beta_fixing} for a 2nd order BCH and first order Magnus expansion. 
	{The optimization step can also be done with reasonable efficiency using a two-step procedure. For a fixed total time $T$, each term $e_q$ is optimized in a serial manner for each step $q$ to find optimal values of time $\tau_q$, and angles $\gamma_q,\beta_q$. These optimal values will depend on the choice of total time $T$, due to the difference in speed and thus magnitude of the counterdiabatic term. This procedure leaves the total number of steps $p$ undefined, and will generally ``overshoot" the total fixed time $T$. As such, in an outer loop the total time $T$ can be optimized through a binary search to find the optimal $T$ for a particular choice of $p$. As $\tau$ changes with $T$, and one should choose a fixed $p$, this optimization is not as immediate as simply setting $T=\sum_q\tau_q$ after one step. In principle, Eq.~\ref{eq:counterdiabatic_QAOA_algorithm} can be optimized simultaneously, but this is usually much slower as a general optimizer is not aware of the specific structure of the function.}
	
	Thus, using minimization, CD-QAOA angles can be derived from some counterdiabatic continuous time protocol to arbitrary orders in perturbation theory. For some fixed $p$, given some counterdiabatic schedule from Eq.~\eqref{eq:counterdiabatic_hamiltonian} and counterdiabatic coefficient $\alpha(\lambda)$ from Eq.~\eqref{eq:alpha_definition}, a set of $2p$ QAOA angles $\{\gamma,\beta\}$ and final counterdiabatic protocol time $T$ can be found by trace minimization.
	
	\subsection{Reverse Counterdiabatic QAOA}\label{sec:reverse_CDQAOA}
	
	The matching procedure can be done in reverse. Given some set of smooth CD-QAOA angles $\{\gamma,\beta\}$, one can minimize Eq.~\eqref{eq:counterdiabatic_QAOA_algorithm}, leaving $\gamma$ and $\beta$ fixed but variationally optimizing the time interval $\tau$, as well as the protocol $\lambda(t)$ and $s(t)$ to some smooth function
	
	\begin{equation}\label{eq:reverse_counterdiabatic_QAOA_algorithm}
	    \underset{\lambda(t),s(t),\tau_q}{\texttt{MIN:}}\quad \sum_q e_q(\gamma_q,\beta_q,\tau_q).
	\end{equation}
	
	The minimization optimizes both the total counterdiabatic annealing time $T=\sum \tau_q$ for given angles $\{\gamma,\beta\}$, as well as generally yields nonlinear counterdiabatic annealing schedules. As will be discussed further in the example of section \ref{sec:3regular_graphs}, optimizing a set of smooth CD-QAOA angles is analogous to optimizing a nonlinear annealing schedule $\lambda(t),s(t)$. To lowest order, the protocol can be derived piecewise by matching a first order Magnus expansion to a second order BCH expansion along each step
	
	\begin{align}\label{eq:reverse_CD-QAOA_protocol}
	    \lambda(t)\quad\Leftrightarrow& \quad \frac{1}{\tau_p}\int_{t_p}^{t_p+\tau_p}\lambda(t) dt= \frac{\gamma_p}{\gamma_p+\beta_p}\quad;\quad \tau_p = \gamma_p+\beta_p\quad;\quad t_p = \sum_{q<p}\tau_p,\\
	    s(t)\quad\Leftrightarrow& \quad \frac{1}{\tau_p}\int_{t_p}^{t_p+\tau_p}s(t) dt= \frac{-\gamma_p\beta_p}{2\tau_p} - \bigg(\frac{\lambda_p - \lambda_{p-1}}{\tau_p}\bigg) \alpha(\lambda_p),\nonumber
	\end{align}
	using a low-order polynomial interpolation. The term $s(t)$ includes removing the counterdiabatic contribution $\alpha(\lambda)$. Not subtracting out the counterdiabatic term will result in an incorrect value of $s(t)$, as protocols with different $p$ and thus total annealing times will result in different magnitudes of $\dot\lambda$ and thus different prefactors of the equivalent 1st order BCH expansion term. If the magnitude $\dot\lambda\ll1$ is small for large $p$ and $T$, the counterdiabatic contribution is small. For intermediate $p$, especially near the edges, excluding the counterdiabatic term will result in incorrect $s(t)$.
	
	\subsection{Conclusion: Counterdiabatic QAOA}
	
	This section has outlined a procedure to map back and forth between continuous-time counterdiabatic evolution, and discrete CD-QAOA angles. Given some counterdiabatic Hamiltonian $H_{CD}$, counterdiabatic protocol $\lambda(t)$, $s(t)$, and number of QAOA steps $p$, one can variationally minimize the trace norm of the difference between continuous time counterdiabatic evolution and effective QAOA evolution to find fixed angles $\{\gamma,\beta\}$, as well as analogous total annealing time $T$
	
	\begin{equation}
	    \big(\;H_{CD}(\lambda,\dot\lambda,s),\;\lambda(t),\;s(t),\;p\;\big)\quad\mapsto\quad \big(\;\{\gamma,\beta\},\;T\;\big).\nonumber
	\end{equation}
	
	Such a mapping always exists for large finite $p$ and $T$ because the sign of the AGP term $\alpha(\lambda)$ is the same as the sign of the second order BCH term. In reverse, given smooth CD-QAOA angles $\{\gamma,\beta\}$, one can variationally minimize the smooth protocol $\lambda,s$ and total annealing time $T$ to best match the effective QAOA evolution to the continuous time counterdiabatic evolution
	
	\begin{equation}
	    \big(\;H_{CD}(\lambda,\dot\lambda,s),\;\{\gamma,\beta\}\;\big)\quad\mapsto\quad  \big(\;\lambda(t),\;s(t),\;T\;\big).\nonumber
	\end{equation}
	
	The next sections will outline three examples of increasing complexity demonstrating this CD-QAOA map. Section \ref{sec:2level_system} will start with an example where the first order AGP is exact, and so $p=1$ CD-QAOA is also exact. Section \ref{sec:ising_chain} will relax the exactness of the AGP with an Ising chain, to demonstrate the varying performance from choosing different counterdiabatic protocols $\lambda(t)$, $s(t)$. Finally, Section \ref{sec:3regular_graphs} will demonstrate the reverse procedure and derive optimal continuous time counterdiabatic protocols from variationally optimized smooth QAOA angles.

	\section{Example 1: 2-level system}\label{sec:2level_system}
	
	\begin{figure}
	    \centering
	    \includegraphics[width=\linewidth]{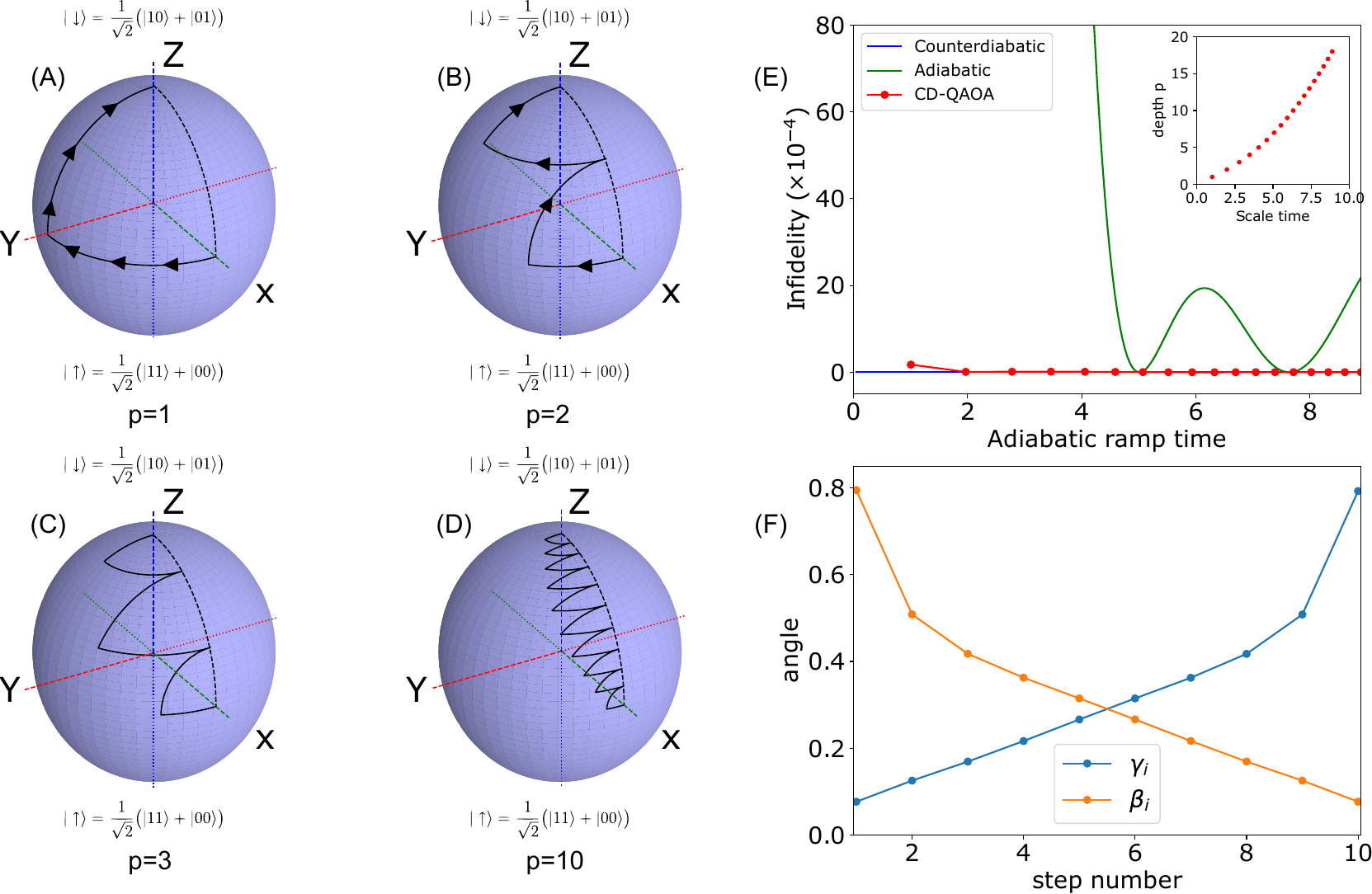}
	    \caption{Counterdiabatic QAOA evolution on the Bloch sphere of the two level system. \textbf{(A-D)} CD-QAOA evolution along each step as a black line evolving continuously along each QAOA unitary. \textbf{(A)} A $p=1$ QAOA may be exact by realizing a rotation of $\pi/2$ around the ``Z" axis, then a rotation of $\pi/2$ around the ``X" axis. \textbf{(B-D)} Higher order CD-QAOA derived from Eq.~\eqref{eq:counterdiabatic_QAOA_algorithm} can mimic the counterdiabatic evolution (black dashed) evolving along the ``XZ" plane by matching the time evolution unitary with the unitary of one step of CD-QAOA using the trace minimization procedure of Eq.~\eqref{eq:counterdiabatic_QAOA_algorithm}. Such a matching is exact up to perturbative error from the finite order of forth-order BCH and third-order Magnus expansions (Red, \textbf{(E)}) and outperforms an equivalent adiabatic-only evolution (Green, \textbf{(E)}). The large-$p$ behavior scales as $T\sim \sqrt{p}$, as expected (inset, \textbf{(E)}). \textbf{(F)} The $p=10$ CD-QAOA angles for a linear schedule $\lambda(t) = t/T$.}
	    \label{fig:2_level_bloch_sphere}
	\end{figure}
	
	Let us begin with the simplest problem: the two level system. We will show that for any $p\geq 1$, we can derive perturbatively exact CD-QAOA angles which reproduce a counterdiabatic evolution. For the two level system, the first order commutator term is the exact adiabatic gauge potential \cite{Chen2010}. Due to this fact, a counterdiabatic Hamiltonian of the form of Eq.~\eqref{eq:counterdiabatic_hamiltonian} can generate exact eigenstates for any protocol speed. In the spirit of choosing Ising-like Hamiltonians for all examples, we choose the two site Ising Hamiltonian
	
	\begin{equation}\label{eq:ham_2level_system}
         H_\text{CD}(\lambda,\dot\lambda) = -\frac{\lambda}{2}\sigma_z^0\sigma_z^1 + (1-\lambda)(\sigma_x^0 + \sigma_x^1) + \frac{i\dot \lambda \alpha}{2}[\sigma_z^0\sigma_z^1,\sigma_x^0 + \sigma_x^1].
    \end{equation}
    \begin{equation}\label{eq:ham_2level_system2}
        H_S = \sigma_x^0 + \sigma_x^1\qquad;\qquad H_T = \frac{-1}{2}\sigma_z^0\sigma_z^1.
    \end{equation}
    
    The Hamiltonian has a $\mathbb Z_2$ symmetry which separates the rank 4 Hilbert space into 2$\times$2 level Hilbert spaces. The maximal state of $H_S$ is the uniform superposition state $|+\rangle$ and is in the $+\mathbb Z_2$ sector spanned by the basis vectors $\{\;(|00\rangle + |11\rangle)/\sqrt{2}\;,\;(|01\rangle + |10\rangle)/\sqrt{2}\;\}$. By symmetry, the two site Ising Hamiltonian with this initial state is a two level system. The variational parameter $\alpha$ can be computed with Eq.~\eqref{eq:alpha_definition} to find
    
    \begin{equation}\label{eq:alpha_singlespin}
        \alpha(\lambda) = \frac{-1}{16(1-\lambda)^2 + \lambda^2}.
    \end{equation}
	
	Evolving $H_\text{CD}$ for any speed results in an exact final eigenstate, as the counterdiabatic term is exact. Similarly, $p=1$ QAOA can generate the exact eigenstate of $\sigma_z^0\sigma_z^1$. The second order BCH term is the adiabatic gauge potential, and higher order terms must be constructed only of terms in Eq.~\eqref{eq:ham_2level_system}, as the operator space is spanned by three operators. This fact can be seen visually by considering the evolution of the two level system on the Bloch sphere in Fig.~\ref{fig:2_level_bloch_sphere}A-D. The initial wavefunction starts in the ``X" direction as an X eigenstate. First, the state is rotated in the ``XY" plane by the $\sigma_z^0\sigma_z^1$ generator by $\pi/2$, eg.~$\gamma=\pi/4$, then rotated in the ``YZ" plane by the $\sigma_x$ generator by $\pi/2$, eg.~$\beta=\pi/4$. The final wavefunction is an exact eigenstate of the $\sigma_z^0\sigma_z^1$ operator, and thus $p=1$ QAOA is exact.

	We may verify this insight by applying the algorithm of Eq.~\eqref{eq:counterdiabatic_QAOA_algorithm} to the two level system for a linear schedule $\lambda(t) = t/T$ and $p=1$ to derive the two CD-QAOA angles. To second order, the Magnus term is
	
	\begin{multline}
	    -i\log\bigg(\mathcal T \exp\bigg(i\int_0^Tdt H_\text{CD}(t)\bigg)\bigg)\approx
	    \bigg(\frac{\pi T}{22} + \frac{T}{2}\bigg)\frac{\sigma_x^0+\sigma_x^1}{2} + \bigg(\frac{\pi T}{22} + \frac{T}{2}\bigg)\sigma_z^0\sigma_z^1 + \bigg(\frac{\pi}{4}-\frac{T^2}{6} \bigg)\frac{\sigma_y^0\sigma_z^1+\sigma_z^0\sigma_y^1}{2}.
	\end{multline}
	
	To third order, the BCH term is
	
	\begin{equation}
	    -i\log\bigg(\exp(i\gamma \sigma_z^0\sigma_z^1)\exp(i\gamma (\sigma_x^0 + \sigma_x^1)/2)\bigg)\approx
	    \bigg( \beta -\frac{\gamma^2\beta}{3}\bigg)\frac{\sigma_x^0+\sigma_x^1}{2}
	    + \bigg( \gamma +\frac{\gamma\beta^2}{3} \bigg)\sigma_z^0\sigma_z^1 +
	    \bigg( \gamma\beta  \bigg)\frac{\sigma_y^0\sigma_z^1+\sigma_z^0\sigma_y^1}{2}.
	\end{equation}
	
	The difference between the Magnus generator and BCH generator may be made exactly zero by minimizing $||\Omega(T) -Z(\gamma,\beta)||$, yielding
	
	\begin{equation}
	    T=0.9974\quad;\quad\gamma=0.2506\pi\quad;\quad\beta = 0.2506\pi.
	\end{equation}
	
	Including higher orders of BCH and Magnus expansion converge the values $\gamma,\beta$ to $\pi/4$. These values show that the perturbative matching of CD-QAOA to a counterdiabatic time evolution can find CD-QAOA angles which are very close to the exact angles, even for small $p$.
	
	\begin{figure}
	    \centering
	    \includegraphics{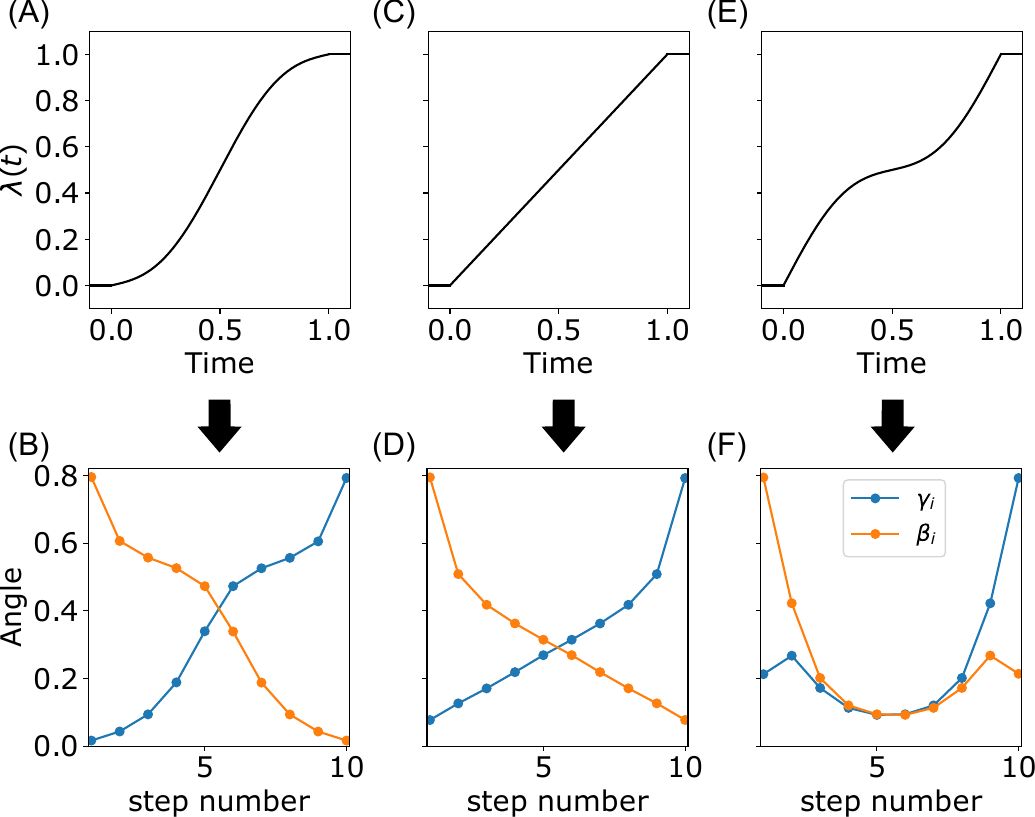}
	    \caption{Induced CD-QAOA angles for a two level system and $p=10$. Depending on the adiabatic schedule $\lambda(t)$ (Top), the induced CD-QAOA angles (bottom) also change. For the two level system, these protocols are all exact up to perturbative corrections. For situations where the adiabatic gauge potential is not exact, different protocols may result in different counterdiabatic performance, and so optimization of smooth QAOA angles is analogous to optimizing an adiabatic schedule $\lambda(t)$.}
	    \label{fig:2level_angles}
	\end{figure}
	
	 An exact higher order $p>1$ set of CD-QAOA angles which mimic a counterdiabatic schedule can also be generated. The exact eigenstate rotates in the ``XZ" plane of the Bloch sphere (black dashed of Fig.~\ref{fig:2_level_bloch_sphere}), while a higher $p$ CD-QAOA would, through repeated rotations around the ``X" and ``Z" axes, set the wavefunction at the end of each step as the exact instantaneous eigenstate along the ``XZ" plane (Fig.~\ref{fig:2_level_bloch_sphere}A-D). Such protocols are not unique, as there is choice of where each step puts the wavefunction in the ``XZ" plane. These non-unique protocols are analogously the choice of adiabatic protocol $\lambda(t)$: different protocols will generate different CD-QAOA angles. Some example adiabatic protocols $\lambda(t)$ and induced CD-QAOA angles for the two level system are shown in Fig.~\ref{fig:2level_angles}. Using a fourth order BCH expansion and third order Magnus expansion to compute the generators $Z$ and $\Omega$, Equation~\eqref{eq:counterdiabatic_QAOA_algorithm} is minimized for various $\lambda(t)$. It is important to reemphasize that such a minimization does not require any quantum simulation, only traces of sums of Pauli operators and computation of Magnus and BCH expansions to constant order.

	\section{Example 2: Ising chain}\label{sec:ising_chain}
	
	A more involved example is the transverse Ising chain, which is a 1d chain of qubits interacting with a nearest neighbor Ising term ($\sigma_z\sigma_z$) and on-site transverse field ($\sigma_x$). We will show that finite $p$ CD-QAOA angles naturally converge to an equivalent finite time counterdiabatic annealing. The ZZ term of the Ising chain is a simple example of a MaxCut objective function on the ring of disagrees if the boundary conditions are periodic.
	
	\begin{equation}\label{eq:isingchain_hamiltonian}
	 H_{CD}(\lambda,\dot\lambda,s) \;=\; \frac{-\lambda}{2}\sum_i\sigma_z^i\sigma_z^{i+1} \;+\; (1-\lambda)\sum\sigma_x^i \;-\; 2\big(s + \dot\lambda \alpha\big)\sum_i\sigma_z^i\sigma_y^{i+1} + \sigma_y^i\sigma_z^{i+1}.
	\end{equation}
	
	\begin{equation}\label{eq:isingchain_hamiltonian2}
	    H_S = \sum\sigma_x^i \qquad;\qquad H_T = \frac{-1}{2}\sum_i\sigma_z^i\sigma_z^{i+1}.
	\end{equation}
	
	The Hamiltonian includes an additional term $\sigma_z\sigma_y$ which is the commutator of the Ising and transverse terms, and acts as the lowest order approximation to the adiabatic gauge potential. The transverse Ising model is exactly solvable (free integrable) using a Jordan Wigner mapping to free fermions \cite{Dziarmaga2005}, and so the eigensystem and nonequilibrium dynamics can be efficiently computed for any system size. There is a critical point at $\lambda=1/2$, $s=0$, with a phase transition between an ordered and disordered phase \cite{Sachdev2011a}, which generates diabatic excitations for any finite speed adiabatic protocol. The system is free integrable, so the adiabatic gauge potential is simple to construct. For $s=0$, the exact AGP is
	
	\begin{align}\label{eq:exact_AGP_TIchain}
	    A(\lambda) =&\alpha_0(\lambda)\bigg(\sum_i \sigma_y^i\sigma_z^{i+1} + \sigma_z^i\sigma_y^{i+1}\bigg) \;+\;\alpha_1(\lambda)\bigg(\sum_i \sigma_y^i\sigma_x^{i+1}\sigma_z^{i+2} + \sigma_z^i\sigma_x^{i+1}\sigma_y^{i+2}\bigg)\nonumber\\
	    +&\alpha_2(\lambda)\bigg(\sum_i \sigma_y^i\sigma_x^{i+1}\sigma_x^{i+2}\sigma_z^{i+3}\ + \sigma_z^i\sigma_x^{i+1}\sigma_x^{i+2}\sigma_y^{i+3}\bigg) + \dots
	\end{align}
	where the ellipsis denote larger Pauli string terms YXX$\cdots$XXZ, for some functions $\alpha(\lambda)$ \cite{KOLODRUBETZ20171}. For $\lambda=0$ or $1$, $\alpha_0=-1/4$ and $\alpha_{i>0}=0$, meaning that the first order counterdiabatic term of $H_{CD}$ is the exact adiabatic gauge potential only at the beginning and end of the annealing process, and a counterdiabatic evolution will not be exact, unlike the two level system of example 1. Different choices of protocol $\lambda(t)$ will generate different final counterdiabatic annealing wavefunctions $|\psi\rangle$ with different amounts of diabatic excitations, and there is some optimal smooth protocol which will minimize this error. The coefficient of the approximate adiabatic gauge potential of the Hamiltonian of Eq.~\eqref{eq:isingchain_hamiltonian} for $s=0$ may be computed using Eq.~\eqref{eq:alpha_definition} as

	\begin{equation}
        \alpha(\lambda) = \frac{-(1-\lambda)^2 -\lambda^2}{8\big((1-\lambda)^2 +\lambda^2 \big)^2 + 8\lambda^2(1-\lambda)^2}.
    \end{equation}
	
\begin{figure}
	    \centering
	    \includegraphics[width=\linewidth]{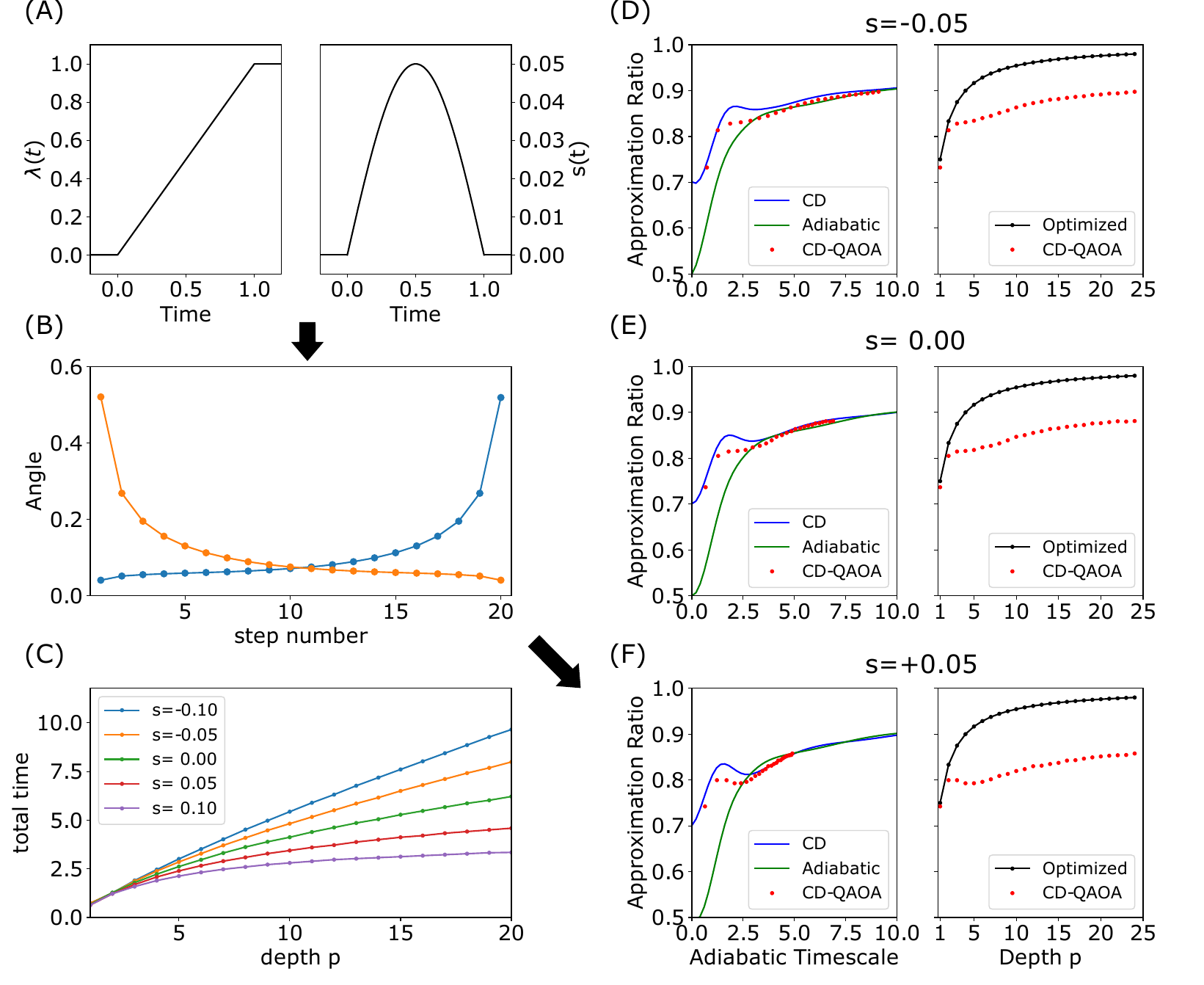}
	    \caption{Counterdiabatic QAOA protocols for the 1d transverse Ising chain. Given a continuous time adiabatic schedule $\lambda(t)$, $s(t)$ \textbf{(A)} and QAOA depth $p$, analogous CD-QAOA angles are derived \textbf{(B)}. \textbf{(D-F)} plots the approximation ratio of the derived CD-QAOA angles (red) for various $p$ and strengths of drive $s$. \textbf{(D-F)} left plots CD-QAOA performance at their analogous adiabatic timescale (red) vs. continuous time counterdiabatic (blue) and adiabatic (green) protocols. In the perturbative limit of $p$ large and $s$ small, the CD-QAOA angles closely follows its analogous counterdiabatic protocol. \textbf{(D-F)} right plots CD-QAOA performance as a function of $p$. Black plots the conjectured maximum of $(2p+1)/(2p+2)$ \cite{farhi2014quantum}, which approach unity as $1/2p$, while CD protocols approach unity as $1/\sqrt{p}$ for $s\neq 0$, suggesting optimal protocols for the Ising chain are non-adiabatic. \textbf{(C)} plots the matched time for a given value of $s$. For $s=0$, the time goes as $\sqrt{p}$, while for $s<0$, time is linear in $p$, in accordance with Eq.~\eqref{eq:p_time_relation}. For $s>0$, there is some maximum total time which cannot be matched for any $p$.}
	    \label{fig:ising_chain_approximation_ratios}
	\end{figure}

	Given some protocol $\lambda(t)$ and $s(t)$ and depth $p$, one may induce the angle parameters $\gamma$ and $\beta$
	
	\begin{equation}
	    \big(\;H_{CD}(\lambda,\dot\lambda,s),\;\lambda(t),\;s(t),\;p\;\big)\quad\mapsto\quad \big(\;\{\gamma,\;\beta\}\;,T\;\big)\nonumber
	\end{equation}
	using the trace minimization procedure of Eq.~\eqref{eq:counterdiabatic_QAOA_algorithm}. For instance, an adiabatic schedule $\lambda(t)=t/T$ and $s(t) = 0.05\sin(\pi t/T)$ (Fig.~\ref{fig:ising_chain_approximation_ratios}A) and $p=20$ derives CD-QAOA angles of Fig.~\ref{fig:ising_chain_approximation_ratios}B with an equivalent continuous annealing timescale of $T=4.588$. We use a 4th order BCH expansion and 3rd order Magnus expansion to compute the effective generators along each step.
	
	\quad
	
	Because the CD-QAOA protocols are directly mapped from continuous adiabatic schedules, it makes sense to compare the CD-QAOA approximation ratio to the continuous time performance via the energy density. Such a comparison is shown in Fig.~\ref{fig:ising_chain_approximation_ratios}D-F for a range of $p$ and values $s(t)=s_0\sin(\pi t/T)$. In the perturbative limit of $p\to\infty$ and $s$ small, the angles $\gamma$ and $\beta$ are also small and so the unitary matching can be made close to exact. As can be seen (eg.~for Fig.~\ref{fig:ising_chain_approximation_ratios}E for $s=0$) as $p$ is large the wavefunction of the QAOA protocol matches that of the analogous continuous time counterdiabatic protocol, indicated by the matching approximation ratio. As $s$ becomes large, higher-order BCH terms accumulate error, diverging the QAOA protocol from its analogous continuous time counterdiabatic protocol.

	For $p=1$, it is notable that the approximation ratio is very close to the optimal value of $C_1=0.7500$. For instance, the $s=0$ $p=1$ induced angles have an approximation ratio of $C_1=0.7368$. Without any query to a quantum simulator, this method is able to compute very close to optimal angles.
	
	As $p$ gets large, the CD-QAOA approaches the continuous counterdiabatic value. It is a natural question to ask how the approximation ratio converges to one as a function of $p$. From Sec.~\ref{sec:METHODS}, for $s=0$ the time of an adiabatic evolution scales as $p\sim \alpha T^2$, where the prefactor is proportional to the average AGP coefficient. For $s\neq 0$, the time of an adiabatic evolution scales as $p\sim s T$, where the prefactor is proportional to the average value of $s$. For continuous adiabatic schedules, the scaling of energy density given a finite-speed passage of a second order phase transition is known from the Kibble-Zurek mechanism \cite{Kolodrubetz2012}.
	Any smooth finite-time protocol
	will cross a symmetry breaking phase transition with some velocity $\dot\lambda_c\sim 1/T$. Around the critical point, the process becomes nonadiabatic and excitations are ``frozen out". The density of these excitations, or ``defects", is related to the correlation length at that freeze-out point. The freeze out time is (roughly) related to the speed of quasiparticle propagation; when the correlation length grows faster than the propagation speed of the quasiparticles, the system is frozen. These excitations, or ``defects", persist through the rest of the adiabatic process on the other side of the critical point. The density of defects means that the final state has a finite energy density. Inverting ground states to maximal states of the objective function, the energy density of defects becomes the approximation ratio as
	
	\begin{equation}
	    C(T) \sim 1 - T^{-d\nu/(1+z\nu)},
	\end{equation}
	where $\nu$ is the critical exponent, and $z$ is the dynamic exponent. The critical and dynamic exponents for the 1d Transverse Ising model are well known \cite{dutta2015quantum} to be $\nu=1$, $z=1$. Given $T\sim p$ for $s\neq 0$, the large $p$ scaling for a CD-QAOA protocol for the ring of disagrees is
	
	\begin{equation}
	    C(p) \sim 1 - p^{-1/2}.
	\end{equation}
	
	The prefactor will depend on the fine details of the protocol, including average value of $s$ and the phase transition crossing speed. If the protocol moves through the critical point in some optimal fashion as a power law \cite{Barankov2008} as $|\lambda-\lambda_c|\sim |t/T|^r$, the approximation ratio instead goes as
	
	\begin{equation}
	    C(T)\sim 1 - (\log(T)/T)^{d/z}\quad\Rightarrow\quad C(p)\sim 1 - p^{-1}\log(p)
	\end{equation}
	
	The extra logarithmic dependence of the approximation ratio is different from the numerically observed optimal values of $C(p)=(2p+1)/(2p+2)$, which has a scaling exponent of $1$ in the large $p$ limit \cite{farhi2014quantum} for the ring of disagrees. Any smooth set of QAOA angles may be induced from some continuous-time adiabatic protocol, and so they must obey the counterdiabatic QAOA scaling of $p^{-1/2}$ or $\log(p)/p$. By exclusion, optimal protocols, which scale as $p^{-1}$, must be non-smooth. Along each step of such a non-smooth protocol, the intermediate wavefunction would be far from any semblance of an instantaneous ground state, and only recovers as an approximate eigenstate at the beginning and end of the QAOA protocol.
	
	The difference between counterdiabatic and optimal QAOA angles for the chain also raises a point about optimizing these angles. For a smooth protocol, one can warm start the optimizer \cite{Egger_2021} with some adiabatic-like values or derived angles from Fig.~\ref{fig:ising_chain_approximation_ratios}. However, for a non-smooth optimal protocol, it is not immediately clear how an optimizer can find optimal angles, which may be hidden within the high-dimensional parameter space. In general, it may be easy to find smooth protocols with adequate QAOA performance derived from adiabatic limits, but it may be hard to find globally optimal protocols with the best performance for fixed $p$ \cite{Day2019}.
	
	Curiously, a $\mu=1$ scaling with no logarithmic correction that is an exactly optimal smooth protocol for the Ising chain has been found in \cite{mbeng2019quantum}. These findings suggest that optimal QAOA protocols derived from optimal passage through critical points may have more subtlety than expected. For example, the counterdiabatic term may suppress just enough excitations to remove any logarithmic correction. Alternatively, the effective auxiliary term $s$ ultimately coherently suppresses excitations of low momentum modes due to the strict locality of QAOA. The presence of such terms may change the low energy behavior of the original Hamiltonian enough to remove the logarithmic correction. Needless to say, a study of such behavior is beyond the scope of this work.
	
	It may be interesting to generalize the exact protocol for the two level system and $p=1$ of Section \ref{sec:2level_system} to the more general Ising chain and larger $p$. For instance, a 2 step BCH expansion for $p=2$ can generate terms which are the next-largest terms in the exact AGP of Eq.~\eqref{eq:exact_AGP_TIchain}, which may enable one to compute exact $p=2$ QAOA angles for a 3 site ring of disagrees. Similarly, going to larger $p$ and using operator Krylov expansions \cite{Viswanath2008} or high order Suzuki-Trotter decomposition \cite{Hatano_2005} to compute the effective QAOA generator, one may be able to compute optimal protocols by matching the generator to the exact counterdiabatic Hamiltonian. In this way, one might be able to analytically derive exact QAOA angles which saturate the conjectured bound of $C(p)=(2p+1)/(2p+2)$ \cite{farhi2014quantum}. Such an investigation is beyond the scope of this work.

	\section{Example 3: regular MaxCut graphs}\label{sec:3regular_graphs}
	 
	As a final example, let us consider QAOA solving MaxCut. We will demonstrate the two-way mapping between counterdiabatic protocols and CD-QAOA angles, and use it to show that transfer of parameters between graphs \cite{brandao2018fixed} and interpolating for larger $p$ \cite{Zhou_2020} are natural consequences. Let us consider a more general Ising Hamiltonian on a graph $\mathcal G$ of the form
	
	\begin{equation}
	    H_\text{CD}(\lambda,\dot\lambda,s) = \lambda \bigg[\frac{1}{2}\sum_{\langle ij\rangle\in \mathcal G}\big(1-\sigma_z^i\sigma_z^j\big)\bigg] \;+\; \big(1-\lambda)\bigg[\sum_{i}\sigma_x\bigg] \;-\; 2\big( s + \dot\lambda \alpha(\lambda)\big)\bigg[\sum_{\langle ij\rangle\in \mathcal G}(\sigma_y^i\sigma_z^j + \sigma_z^i\sigma_y^j)\bigg],
	\end{equation}
	
	\begin{equation}
	    H_S = \sum_{i}\sigma_x\qquad;\qquad H_T = \frac{1}{2}\sum_{\langle ij\rangle\in \mathcal G}\big(1-\sigma_z^i\sigma_z^j\big).
	\end{equation}
	
	Maximal energy states of the target Hamiltonian encode MaxCut solutions for the graph $\mathcal G$, and the maximal state of the simple Hamiltonian is a uniform superposition state $|+\rangle$. The third term is the commutator of the first two terms, and serves as the lowest order approximate adiabatic gauge potential, as well as auxiliary field $s$. The Hamiltonian can include example 2 as a special case if the graph $\mathcal G$ is an infinite chain of singly connected qubits. This example will focus on 3 regular graphs only.

\begin{figure}
	    \centering
	    \includegraphics[width=\linewidth]{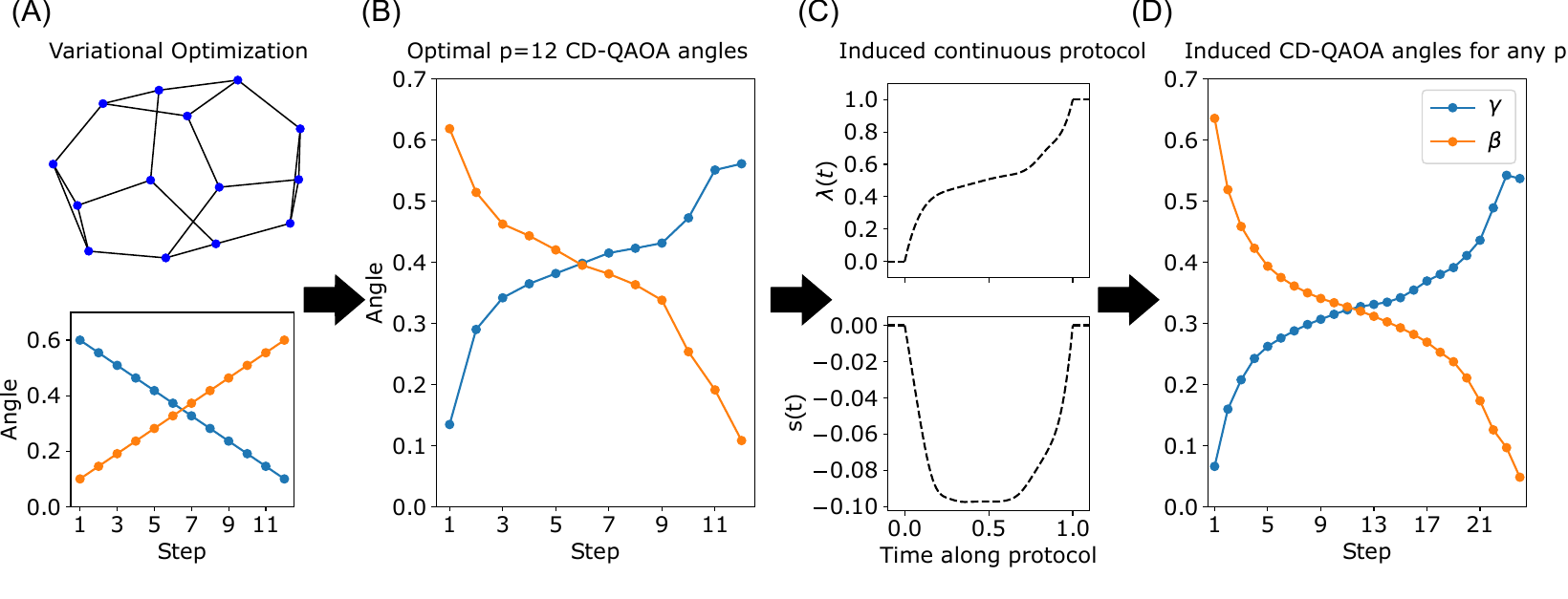}
	    \caption{Inducing continuous adiabatic protocols and CD-QAOA angles from variational optimization. Given some small graph instance and adiabatic-inspired initial point \textbf{(A)}, a classical optimizer maximizes the objective function and yields an optimal set of smooth CD-QAOA angles \textbf{(B)}. Given these angles and a variationally derived prefactor for the adiabatic gauge potential, a continuous protocol for $\lambda(t)$ and $s(t)$ can be induced \textbf{(C)}. With the continuous protocol, a set of CD-QAOA angles for any $p$ is induced \textbf{(D)}.} 
	    \label{fig:3r_derived_protocol}
	\end{figure}

	Similar to the first two examples, given an counterdiabatic annealing schedule $\lambda(t)$ and $s(t)$, one can compute a set of CD-QAOA angles which mimics the continuous time evolution. For this example, we show the opposite approach. A set of smooth, optimized CD-QAOA angles can induce the continuous time counterdiabatic annealing schedule which best mimics the CD-QAOA angles, using the methods of Sec.~\ref{sec:reverse_CDQAOA}. This procedure is shown in Fig.~\ref{fig:3r_derived_protocol}.
	
	First, a classical optimizer maximizes the objective function (Fig.~\ref{fig:3r_derived_protocol}A) for a specific graph $\mathcal G$ and $p$, and yield a set of angles $\{\gamma,\beta\}$ (Fig.~\ref{fig:3r_derived_protocol}B). If reasonably smooth, the angles can be used to induce the continuous time counterdiabatic protocol using Eq.~\eqref{eq:reverse_CD-QAOA_protocol} (Fig.~\ref{fig:3r_derived_protocol}C). It is important to subtract the counterdiabatic term. In the very large $p$ limit, the total time is large and thus the counterdiabatic term is small and thus has negligible change for the form of $s(t)$. However, for intermediate $p$, different values of $p$ would spuriously change the form of $s(t)$ if the AGP value was not excluded due to a small adiabatic timescale and thus large $\dot\lambda\alpha$.  For a $\nu$-regular graph with no cycles of size 3, the variational parameter can be computed using Eq.~\eqref{eq:alpha_definition} to be
	
	\begin{equation}
        \alpha(\lambda) = \frac{-32(1-\lambda)^2 -8(3\nu-2)\lambda^2}{256\big((1-\lambda)^2 + 4(3\nu-2)\lambda^2)\big)^2 + 256\lambda^2(1-\lambda)^2(\nu-1) + 96(\nu-1)(\nu-2)\lambda^4}.
    \end{equation}

	The annealing schedule $\lambda(t)$, $s(t)$ can then be induced using Eq.~\eqref{eq:reverse_CD-QAOA_protocol} using a polynomial interpolation (Fig.~\ref{fig:3r_derived_protocol}C). Then, given the fixed smooth protocol, the process can be done in reverse to derive new counterdiabatic QAOA angles (Fig.~\ref{fig:3r_derived_protocol}D) for a different value of $p$.
	
	Optimizing a counterdiabatic QAOA schedule is equivalent to optimizing a smooth protocol $(\lambda,s)$. Different continuous time protocols have better or worse performance, and for some constraint of total time $T$ and smooth ansatz $\lambda(t),s(t)$, there must be some smooth continuous time protocol that is optimal. Note that without the restriction of smoothness, by Pontryagin’s minimum principle optimal protocols are bang-bang, eg.~QAOA \cite{Yang_2017,Brady2021}. In the large $p$ limit, we know from section \ref{sec:METHODS} that a counterdiabatic QAOA closely reproduces a continuous adiabatic schedule and, from Eq.~\eqref{eq:reverse_CD-QAOA_protocol}, one can be mapped to the other. In this way, the classical optimizer is effectively searching for the best smooth adiabatic schedule \cite{schiffer2021adiabatic} by optimizing angles which best mimic that schedule.
	
	We comment that the best continuous schedule may differ slightly from that derived from CD-QAOA angles, due to a slight bias. For a continuous protocol, the total time $T$ is fixed, while for a CD-QAOA protocol the number of steps $p$ is fixed, and the time depends on the average value of $s$, as $T\sim s p$. By the adiabatic theorem, larger total evolution times have less excitations, and thus the optimizer is biased to a larger average value of $s$.
	
	\quad
	
	We show an example of this implementation in Fig.~\ref{fig:3r_derived_protocol}. Here, variational parameters for a $p=12$ QAOA are initialized in an adiabatic inspired manner for a small 3 regular graph with 14 vertices (Fig.~\ref{fig:3r_derived_protocol}A). These initial values are optimized for the particular graph using a numerical gradient ascent procedure, which returns only after settling into a local minimum where the gradient is zero. The resultant optimal angles (Fig.~\ref{fig:3r_derived_protocol}B) are smooth and mimic a continuous time counterdiabatic schedule, as expected. The approximation ratio is $C_{12}=0.99004$, very close to the exact result of unity. Using the procedure of Eq.~\eqref{eq:reverse_CD-QAOA_protocol} and a polynomial interpolation, the analogous continuous counterdiabatic protocol is derived, with an equivalent adiabatic time of $T\approx 9.3$ (Fig.~\ref{fig:3r_derived_protocol}C). With this continuous protocol, an equivalent CD-QAOA protocol may be derived for any value of $p$ using the variational minimization procedure of Eq.~\eqref{eq:counterdiabatic_QAOA_algorithm}. For example, this protocol derives $p=24$ angles (Fig.~\ref{fig:3r_derived_protocol}D) with an approximation ratio $C_{24}=0.99994$ and an equivalent adiabatic time of $T\approx 15.4$.
	
	The procedure of inducing larger $p$ angles from some fixed $p$ optimization formalizes the original observations of \cite{Zhou_2020}, which interpolate smooth QAOA angles for a certain $p$ to induce different angles $p'$ with good performance. These observations can now be justified under these counterdiabatic procedures: both $p$ and $p'$ share the same underlying approximately optimal counterdiabatic protocol. However, this procedure also allows for computing CD-QAOA angles for small $p\sim 1$, where the effective annealing timescale is small and the contribution from the counterdiabatic term $\dot\lambda \alpha(t)$ may be large. Near the edges of an evolution where the rate of change is large, the counterdiabatic term $\dot\lambda \alpha(\lambda)$ may be large, even for $p$ large, a phenomenon that a simple interpolation may fail to capture.
	
	It has has been numerically observed that optimal QAOA angles of one graph can be transferred to other graphs of the same class~ \cite{brandao2018fixed}. It is reasonable to expect that the optimal continuous protocol derived from CD-QAOA angles of one graph should be transferable.	If two graphs share the same low energy eigenspectrum, an optimal counterdiabatic protocol for one graph should be similar to an optimal protocol $\lambda(t)$, $s(t)$ for the other.	
	
	By extension, it seems reasonable to expect that optimized CD-QAOA angles for one graph should have a similar performance to the other. The counterdiabatic protocol induced from the optimized for the first graph should be close to optimal for the other graph. Similarly, the CD-QAOA angles induced from that protocol for the new graph should also be similar to those for the old graph. In lieu of this procedure, which ultimately results in very similar CD-QAOA angles, one can instead directly transfer angles from one graph to another. This indicates that transfer of parameters, at least for smooth sets of angles, is a natural consequence of counterdiabaticity.

	\section{QAOA is better than continuous-time adiabatic protocols}\label{sec:QAOA_is_better_than_adiabatic}
	
	A natural question to ask is how continuous time adiabatic annealing protocols perform in comparison to QAOA. Here we show that under specific restrictions QAOA is better, by comparing finite time adiabatic annealing with finite time counterdiabatic annealing via a chain of inequalities relating approximation ratios:
	
	\begin{equation}
	\begin{bmatrix}
	    \text{\scriptsize{Optimal QAOA}}\\
	    \text{\scriptsize{with fixed $T$}}\\(\;1\;)
	\end{bmatrix}
	 \;\geq\;
	 \begin{bmatrix}
	    \text{\scriptsize{Optimal CD-QAOA}}\\
	    \text{\scriptsize{with fixed $T$}}\\(\;2\;)
	\end{bmatrix}
	 \;\geq\;
	 \begin{bmatrix}
	    \text{\scriptsize{CD-QAOA}}\\
	    \text{\scriptsize{with fixed $T$, $p$}}\\(\;3\;)
	\end{bmatrix}
	\;\underset{p\text{ large}}{=}\;
	 \begin{bmatrix}
	    \text{\scriptsize{Counterdiabatic}}\\
	    \text{\scriptsize{with fixed $T$}}\\(\;4\;)
	\end{bmatrix}
	\;\geq\;
	\begin{bmatrix}
	    \text{\scriptsize{Adiabatic}}\\
	    \text{\scriptsize{with fixed $T$}}\\(\;5\;)
	\end{bmatrix}.\nonumber
	\end{equation}

	Suppose for some optimization problem, the $2p$ optimal QAOA angles (1) with a total magnitude $T$, which we call an ``angle budget"
	
	\begin{equation}
	    T = \sum_{q=1}^p |\gamma_q| + |\beta_q|.
	\end{equation}
	
	Now, suppose this angle budget $T$ acts as an optimization constraint. While the constraint is not as useful for digital computing, it has a natural interpretation as the total adiabatic evolution time, which is a common constraint for QAA. By the nature of global optima under constraints, these optimal angles (which may be non-smooth) have a greater than or equal approximation ratio than any CD-QAOA angles (2) with fixed angle budget $T$ and variable $p$. This includes optimal CD-QAOA angles which optimize the effective counterdiabatic schedule $\lambda(t)$ and some $p(T)$ to best match counterdiabatic terms. Because $\alpha(\lambda)$ has the same sign as the second order BCH term, $p(T)$ will always be finite.
	
	The value $p$ can always be made large and fixed (3), which again by nature of optima under constraints must have a less than or equal approximation ratio to the optimal CD-QAOA angles. Increasing $p$ reduces the strength of the effective counterdiabatic term; If large enough, the higher order BCH error can be made small enough to mimic the effective Hamiltonian (3) by suppressing higher-order error terms
	
	\begin{equation}\label{eq:Heff_sec8}
	    H_\text{eff} = \frac{\gamma(t)}{\gamma + \beta} H_T + \frac{\beta(t)}{\gamma + \beta} H_S - \frac{i\gamma\beta}{\gamma + \beta} [H_T,H_S]
	\end{equation}
	with some (less than optimal) effective counterdiabatic term $-\gamma_q\beta_q/(\gamma_q+\beta_q)$. Finally, due to the sign of the effective CD term (See Sec.~\ref{sec:shortcuts_to_adiabaticity}), the finite time counterdiabatic annealing (4) has a larger approximation ratio than an evolution of the adiabatic-only annealing (5). By transitivity, any optimal QAOA with angles $\{\gamma,\beta\}$ has a better approximation ratio than an adiabatic-only finite time annealing evolution with total time $T=\sum_q|\gamma_q|+|\beta_q|$. In this sense, QAOA can always outperform an adiabatic evolution by at least reproducing a counterdiabatic evolution.

	This observation is an instance of Pontryagin’s minimum principle \cite{Yang_2017,Brady2021}: given some fixed total time for adiabatic evolution, optimal protocols are bang-bang, as long as the controls are nonsingular \cite{Seraph2018}. However, the critical intermediate step of CD-QAOA protocols requires observation of the \textit{sign} of the counterdiabatic term. For the lowest order approximation of the adiabatic gauge potential, $\alpha(\lambda)\leq 0$ always by Eq.~\eqref{eq:alpha_definition}. Similarly, the second order BCH term is also $<0$ always. The matching of the effective CD-QAOA Hamiltonian to the counterdiabatic Hamiltonian with finite $p$ and fixed $T$ is only possible with this choice of signs: for some small but negative $\alpha$, the effective Hamiltonian of eq.~\eqref{eq:Heff_sec8} will have better adiabatic performance on the ground or maximal state of the model. If $\alpha>0$ the variational matching procedure would choose $p\to\infty$ to minimize the weight of the commutator term to be optimal. Due to this, the last inequalities $(3)\geq (5)$ hold and QAOA does better than adiabatic, as QAOA at least reproduces a counterdiabatic evolution instead of an adiabatic one.
	
	Although we show that smooth CD-QAOA can have good finite-$p$ performance, it is not necessarily clear that smooth angles are always optimal. By Pontryagin’s minimum principle, optimal protocols are QAOA-like, but there is no necessary requirement on smoothness. An optimal protocol for a given $p$ may be highly non-adiabatic, with the intermediate state being far from an instantaneous ground state and angles being highly nontrivial. It may be an interesting future study to find when optimal QAOA angles are smooth, or non-smooth, which may motivate an optimizer which biases its search to ``non-smooth" regions of parameter space.
	
	\section{Conclusion}\label{sec:conclusion}
	
	In this work, we have made the connection between QAA and QAOA explicit by inspecting the regime of $p$ large but finite. The results are centered around one observation: the second order BCH term, which determines the effective QAOA evolution generator, is the same as a low order approximation of the adiabatic gauge potential, which suppresses diabatic excitations.
	By identifying and matching counterdiabatic annealing with QAOA evolution, we analytically construct sets of CD-QAOA angles which best mimic the continuous time evolution.
	Importantly, this process is completely tractable for any system size. Computation of coefficients requires only well-behaved variational minimization of traces of Pauli operators, which can be done efficiently on a classical computer without needing quantum simulations.
	While these results are perturbatively exact in the large-$p$ limit, the CD-QAOA angle derivation is well behaved and has good performance even for small to intermediate $p\lesssim 10$. This suggests that these analytically derived angles can be used directly \cite{Streif_2020} bypassing the optimization step, or as variational warm starts on tomorrow's NISQ devices.
	
	{While this method was applied to the MaxCut problem, these counterdiabatic insights may be applied to a wide range of objective Hamiltonians. For example, these methods may be applied to the variational quantum eigensolver to find low energy states of quantum chemistry Hamiltonians, which have off-diagonal matrix elements and are thus more general than binary optimization problems. Given some general setting of simple and target Hamiltonians $H_S$ and $H_T$, a QAOA-like protocol can generally be derived using these methods to prepare low energy states in a counterdiabatic manner. Such a study is beyond the scope of this paper, although these results should transfer naturally to off-diagonal Hamiltonians.}

	Critical to the matching procedure is the fact that the sign of the variational adiabatic gauge potential coefficient $\alpha$ is the same sign as the second order BCH coefficient. This implies that for large $T$, there is some large but finite $p(T)$ which is optimal given that counterdiabatic evolution has better performance than adiabatic-only annealing. The existence of such finite-$p$ protocols also implies that, in some sense, QAOA does better than finite time adiabatic-only annealing. Given some time $T$ or equivalently angle budget $T=\sum_q|\gamma_q|+|\beta_q|$, the effective QAOA evolution can include terms which reduce diabatic excitations, or in accordance with Pontryagin’s minimum principle \cite{Yang_2017} take nonadiabatic paths with better use of the angle budget.
	
	Matching of $p$ with $T$, as well as improving performance over adiabatic behavior, also allows for insight of lower bounds on how the approximation ratio must converge with $p$. Using Kibble Zurek scaling, we show that the approximation ratio converges to one at least polynomially $1-C(p)\sim p^{-\mu}$, where exponent $\mu$ is derived from the critical exponents of the model. While this bound exists for smooth CD-QAOA angles, optimal angles may converge faster than the adiabatic limit. For instance, we find exponent $\mu=1/2$ for the transverse Ising model, slower than the conjectured optimal behavior of $\mu=1$. This suggests that optimal QAOA angles may be non-smooth and potentially hard to find.
	
	Another important part of the construction is the addition of the auxiliary term $s$, which adds extra terms to the effective QAOA Hamiltonian. Instead of following the ground state of the simple Hamiltonian $\lambda H_S + (1-\lambda)H_T$, the QAOA wavefunction may follow the ground state of a Hamiltonian with extra terms $H(t) + s(t)\tilde H(t)$. We numerically observe that near the edges of the optimized protocols of example 3, the protocol is ``fast", and thus counterdiabatic, whereas near the middle, the protocol is ``slow" and thus adiabatic only, except with a large term $s$ which allows for a slow effective annealing time $T\sim p$. The necessity of the auxiliary term $s$ to match optimized QAOA angles with an effective QAOA Hamiltonian suggests that further study is required on the subject beyond the scope of this work.
	
	Note that optimal QAOA angles may be degenerate (see for example Table 1 of \cite{wurtz2021}). For example, the approximation ratio is unchanged under $\{\gamma,\beta\}\to\{-\gamma,-\beta\}$ under time reversal symmetry. Additionally, a $U(2)$ symmetry allows addition of factors of $\pi$ to angles. More non-trivial symmetries may exist, which relate points in parameter space under symmetry transformations. A set of smooth CD-QAOA angles could be ``scrambled" under these symmetry transformations into a class of non-smooth QAOA angles that are nonetheless counterdiabatic. Alternatively, a non-smooth set of QAOA angles may be in fact non-counterdiabatic if they are not part of a class that cannot be made smooth under symmetry transformations. It may be interesting to distinguish between optimal angles which are counterdiabatic under transformations, and those which are not; however, this is beyond the scope of this work.
	
	We also show that two heuristically observed behaviors of QAOA are natural consequences of CD-QAOA. It has been observed \cite{brandao2018fixed} that optimal parameters for one MaxCut graph ``transfer", or have similar performance, on other graphs with a similar low-energy eigenspectrum. We posit that this is a consequence of the fact that smooth CD-QAOA angles can derive continuous-time counterdiabatic annealing schedules $\lambda(t)$, and vice versa. Similarly, the interpolation procedure of \cite{Zhou_2020}, which induces parameters for level $p$ given optimal parameters of $p-1$, can be justified in the same way. We add the extra improvement of including the counterdiabatic term to adjust to different annealing times from different $p$.

	In conclusion, this paper studies the subtle interplay between QAOA and the quantum adiabatic algorithm (QAA). Ultimately, QAOA is always at least \textit{counterdiabatic}, not just \textit{adiabatic}. It is an interesting future direction to further explore the extra advantage of QAOA and other digital algorithms over their analog counterparts. Here, we show that QAOA can at least beat quantum annealing, though give no bounds on improvement. Given an analogous continuous counterpart, it will be interesting to see where and how there is an explicit separation in performance given analogous quantum resources.
	
	\section*{Acknowledgements}
	
	This work was supported by NSF STAQ project (PHY-1818914) and the Defense Advanced Research Projects Agency (DARPA) under Contract No. HR001120C0068.
	
	\appendix
	
    \section{Appendix: BCH and Magnus Expansion terms}
    
    For completeness, we include the Baker-Campbell-Hausdorff and Magnus expansions here. The BCH expansion is a sum of terms $Z \;= Z_1 + Z_2 + Z_3 + Z_4+\dots$ where
    
    \begin{align}\label{eq:BCH_expansion}
	    Z_1&=\gamma H_T + \beta H_S\\
	    Z_2&=-\frac{i\gamma\beta}{2}[H_T,H_S]\nonumber\\
	    Z_3&=  \frac{\gamma\beta^2}{12}\big[H_S,[H_T,H_S]\big] -\frac{\gamma^2\beta}{12}\big[ H_T,[ H_T,H_S]\big]\nonumber\\
	    Z_4&=\frac{-i\gamma^2\beta^2}{24}\big[H_T,\big[H_S,[H_T,H_S]\big]\big]\nonumber\\
	    Z_5&=\dots\nonumber
	\end{align}
	
	Higher order terms can be found constructively, eg.~\cite{Hatano_2005}. The Magnus expansion is sum of terms $\Omega \;= \Omega_1 + \Omega_2 + \Omega_3+\dots$ where, for a smooth protocol around $t_0$ is $\lambda(t-t_0)\approx \lambda_0 + \dot \lambda t + \ddot \lambda t^2/2$,
	
    \begin{align}\label{eq:magnus1}
        \Omega_1 &= \int_{t_0}^{t_0+\tau}dt\lambda(t) H_T + (1-\lambda(t)) H_S + i\big(s(t) + \dot \lambda(t) \alpha(\lambda(t))\big)[H_T,H_S]\nonumber\\
        &=\tau \overline \lambda H_T  + \tau (1-\overline \lambda ) H_S +i\tau\big( \overline s + \dot\lambda \overline{\alpha}\big) [H_T,H_S],\\\quad\nonumber
    \end{align}
    where $\overline \lambda = \frac{1}{\tau}\int_{t_0}^{t_0+\tau}dt\lambda(t)$ and $\overline{\overline \alpha}=\int_{\lambda(t_0)}^{\lambda(t_0+\tau)}d\lambda\alpha(\lambda)$ are the average coefficients over the interval. For smooth $\alpha(\lambda)$ and adiabatic schedule $\lambda(t)$, $\overline{\overline \alpha} = \dot \lambda \tau \overline \alpha$, with $\overline\alpha=\frac{1}{\tau} \int_{t_0}^{t_0+\tau} dt \alpha(t)$ the average coefficient over the interval. The second order term is

    \begin{align}\label{eq:magnus2}
        \Omega_2&=-\frac{1}{2}\int_{t_0}^{t_0+\tau}dt_1 \int_{t_0}^{t_1}dt_2 [H_\text{CD}(t_1), H_\text{CD}(t_2)]\\
        &=\frac{1}{2}\int_{t_0}^{t_0+\tau}dt_1 \int_{t_0}^{t_1}dt_2\bigg(\lambda(t_1)\big(1-\lambda(t_2)\big) - \big(1-\lambda(t_1)\big)\lambda(t_2)\bigg)[H_T,H_S] \\
        &\qquad\qquad\qquad\qquad\quad+\bigg(\lambda(t_1)\dot\lambda(t_2)\alpha(t_2) - \lambda(t_2)\dot\lambda(t_1)\alpha(t_1)\bigg)[H_T,[H_T,H_S]]\nonumber\\
        &\qquad\qquad\qquad\qquad\quad+\bigg((1-\lambda(t_1))\dot\lambda(t_2)\alpha(t_2) - (1-\lambda(t_2))\dot\lambda(t_1)\alpha(t_1)\bigg)[H_S,[H_T,H_S]]\nonumber\\
        &\approx \Bigg(\frac{-\dot\lambda \tau^3}{12} + \frac{-\ddot \lambda \tau^4}{24}\Bigg)[H_T,H_S] + \mathcal O(\dot \lambda^2\tau^3,\ddot \lambda\tau^3,\dot \lambda \ddot \lambda \tau^4,\dots)
    \end{align}
    %
    %
    %
    where the higher order contributions are from the doubly nested commutators arising from the counterdiabatic term. The third order term $\Omega_3$ will be constructed of at least doubly nested commutators such as $[H_T,[H_T,H_S]]$ at lowest order, and have prefactors such as $\dot \lambda \tau^4$, $\ddot \lambda \tau^5$, $\dot \lambda^2\tau^5$. The terms of third order and higher can be found in \cite{Blanes_2009}.
	
	\section{Derivation of Eq.~\ref{eq:alpha_singlespin}}
	
	In this section, we derive Eq.~\ref{eq:alpha_singlespin} in more detail. This has been done in several places elsewhere, e.g.~ref.~\cite{KOLODRUBETZ20171} but is repeated for completeness here. The AGP term is defined by Eq.~\ref{eq:alpha_definition}
	
	\begin{equation}
	    A(\lambda) = i\alpha(\lambda) [H_T,H_S]\quad\text{with}\quad \alpha(\lambda) = \frac{\Big|\Big|[H,\partial_\lambda H]\Big|\Big|_\PP}{\Big|\Big|\big[[H,\partial_\lambda H],H\big]\Big|\Big|_\PP}.
	\end{equation}
	
	and the Hamiltonian terms for the 2 site Ising model are defined by Eq.~\ref{eq:ham_2level_system2}
	
	\begin{equation}
        H_S = \sigma_x^0 + \sigma_x^1\qquad;\qquad H_T = \frac{-1}{2}\sigma_z^0\sigma_z^1.
    \end{equation}
	
	Plugging in the derivative of $\partial_\lambda H=-\sigma_z^0\sigma_z^1/2 - \sigma_x^0-\sigma_z^1$, the commutators are
	
	\begin{align}
	    \big [H,\partial_\lambda H\big]&=\big[-\frac{\lambda}{2}\sigma_z^0\sigma_z^1/2 + (1-\lambda)(\sigma_x^0 + \sigma_x^1)\;,\;-\sigma_z^0\sigma_z^1 - \sigma_x^0-\sigma_z^1\big ],\\
	    &=(+i\lambda + i(1-\lambda))(\sigma_y^0\sigma_z^1 + \sigma_z^0\sigma_y^1) = i(\sigma_y^0\sigma_z^1 + \sigma_z^0\sigma_y^1),\\
	    \big [ [H,\partial_\lambda H],H\big] &= i\big[\sigma_y^0\sigma_z^1 + \sigma_z^0\sigma_y^1\;,\;-\frac{\lambda}{2}\sigma_z^0\sigma_z^1/2 + (1-\lambda)(\sigma_x^0 + \sigma_x^1)\big],\\
	    &=\lambda(\sigma_x^0 +\sigma_x^1) + 4(1-\lambda)(\sigma_z^0\sigma_z^1 - \sigma_y^0\sigma_y^1).
	\end{align}
	
	The norm is defined as $||A||_{\mathcal P}=\tr{\mathcal P A^2}$. The projective subspace are the $\mathbbm{Z}_2$ symmetric states $\{(|11\rangle + |00\rangle)/\sqrt{2}\;,\;(|01\rangle+|10\rangle)/\sqrt{2}\}$. The square of the terms are
	
	\begin{align}
	    \big([H,\partial_\lambda H]\big)^2 &= -2(\mathbbm 1+\sigma_x^0\sigma_x^1),\\
	    \big(\big [ [H,\partial_\lambda H],H\big]\big)^2 &= 2(\mathbbm 1+\sigma_x^0\sigma_z^1)(\lambda^2 + 16(1-\lambda)^2).
	\end{align}
	
	Observe that the operators select the $+1$ sector of $\mathbbm Z_2$, as $\sigma_x^0\sigma_x^1$ is the generator of $\mathbbm Z_2$ symmetry, so the trace over the subspace and operators is 4 for both numerator and denominator. Eq.~\eqref{eq:alpha_singlespin} can then be found by dividing the numerator by denominator above. A similar procedure can be done for the following examples, keeping care to factors of $-1$, $2$, and $i$.
	
\normalem
	\bibliographystyle{apsrev4-1}
	\bibliography{citationlist} 
	
\end{document}